\newcommand{\al}{\alpha}

\newcommand{\be}{\beta}

\newcommand{\de}{\delta}

\newcommand{\la}{\lambda}

\newcommand{\si}{\sigma}

\newcommand{\beena}{\begin{eqnarray}}
\newcommand{\eena}{\end{eqnarray}}

\newcommand{\tn}{\textnormal}

\newcommand{\deriv}[2]{\frac{\partial{#1}}{\partial{#2}}}

\newcommand{\PPi}{\boldsymbol{\Uppi}}

\newcommand{\bb}{\mathcal{B}}
\newcommand{\ff}{\mathbf{F}}
\newcommand{\XX}{\textbf{X}}
\newcommand{\TT}{\textbf{T}}
\newcommand{\xx}{\textbf{x}}

\newcommand{\m}{\textbf{m}}
\newcommand{\EE}{\textbf{E}}
\newcommand{\ee}{\textbf{e}}
\newcommand{\M}{\textbf{M}}

\newcommand{\C}{\mathbf{C}}
\newcommand{\B}{\mathbf{B}}
\newcommand{\CS}{\boldsymbol{\sigma}}

\pdfoutput=1

\documentclass{article}%%%%where rsproca is the template name

\usepackage{upgreek}
\usepackage{authblk}
\usepackage{geometry}
\usepackage{cite}
\usepackage{hyperref}
\usepackage{fancyhdr}
\usepackage[toc,page]{appendix}
\usepackage{pdfpages}
\usepackage{todonotes}
\usepackage{subfigure}
\usepackage{bigints}
\usepackage{graphicx}
\usepackage{amssymb}
\usepackage{amsmath}
\usepackage[infoshow,debugshow]{tabularx}
\usepackage{relsize}
\usepackage{color}
\usepackage{cleveref}
\usepackage{indentfirst}
\usepackage{setspace}
\usepackage[english]{babel}
\usepackage{mathtools}
\usepackage{nicefrac}
\DeclareMathOperator{\tr}{tr}
\DeclareMathOperator{\cc}{C}

\DeclareMathOperator{\Grad}{Grad}

\DeclareMathOperator{\diag}{diag}

\DeclareMathOperator{\tra}{T}

\DeclareMathOperator{\el}{e}
\DeclareMathOperator{\Lo}{L}
\DeclareMathOperator{\A}{A}
\DeclareMathOperator{\T}{T}
\DeclareMathOperator{\iso}{ISO}
\DeclareMathOperator{\MR}{MR}

\begin{document}

\title{\textbf{A modified formulation of quasi-linear viscoelasticity for transversely isotropic materials under finite deformation}}

\author[1]{Valentina Balbi}
\author[2,3]{Tom Shearer}
\author[2]{William J Parnell}
\affil[1]{\small{School of Mathematics, Statistics and Applied Mathematics, NUI Galway, University Road, Galway, Ireland}}
\affil[2]{\small{School of Mathematics, University of Manchester, Oxford Road, Manchester M13 9PL, United Kingdom}}
\affil[3]{\small{School of Materials, University of Manchester, Oxford Road, Manchester M13 9PL, United Kingdom}}
\date{}

%\address{1.School of Mathematics, Statistics and Applied Mathematics, NUI Galway, University Road, Galway, Ireland\\
%2.School of Mathematics, University of Manchester, Oxford Road, Manchester M13 9PL, United Kingdom\\
%3.School of Materials, University of Manchester, Oxford Road, Manchester M13 9PL, United Kingdom}

%\subject{Applied Mathematics, Applied Physics, Mechanics}

%\keywords{}

%\corres{V.\ Balbi \\  \email{v.balbiv@gmail.com}}

%%%% Abstract text to be placed here %%%%%%%%%%%%

\maketitle
\begin{abstract}
The theory of quasi-linear viscoelasticity (QLV) is modified and developed for transversely isotropic (TI) materials under finite deformation. For the first time, distinct relaxation responses are incorporated into an integral formulation of nonlinear viscoelasticity, according to the physical mode of deformation. The theory is consistent with linear viscoelasticity in the small strain limit and makes use of relaxation functions that can be determined from small-strain experiments, given the time/deformation separability assumption. After considering the general constitutive form applicable to compressible materials, attention is restricted to incompressible media. This enables a compact form for the constitutive relation to be derived, which is used to illustrate the behaviour of the model under three key deformations: uniaxial extension, transverse shear and longitudinal shear. Finally, it is demonstrated that the Poynting effect is present in transversely isotropic, neo-Hookean, modified QLV materials under transverse shear, in contrast to neo-Hookean \textit{elastic} materials subjected to the same deformation. Its presence is explained by the anisotropic relaxation response of the medium.
\end{abstract}
%%%%%%%%%%%%%%%%%%%%%%%%%%%

%\begin{fmtext}

\section{Introduction}
The ability to predict time-dependent deformation in soft, compliant solids is important when modelling a diverse range of materials, such as reinforced polymers, elastomers and rubbers, and is of increasing importance in the context of soft tissue mechanics. In many of these applications, the microstructure of the medium in question dictates that the material response is both strongly anisotropic and viscoelastic. Additionally, given the compliant nature of these materials, it is necessary to accommodate finite strains. The field of finite strain nonlinear viscoelasticity has a rich history and a huge range of alternative constitutive forms has been proposed. The review by Wineman \cite{Win-09} provides a comprehensive overview of the current state-of-the-art of the field. In this introduction, we provide a summary of some details of existing models in order to provide context and to motivate the present study, specifically with respect to the study of viscoelastic anisotropy.

In the most general viscoelastic setting, the constitutive relation relating stress to deformation has the Cauchy stress $\mathbf{T}(t)$, where $t$ is time, written in the general form \cite{Win-09}
\begin{align}
\mathbf{T}(t) &=  \mathbf{F}(t)\mathcal{G}\left[\mathbf{C}(t-s)\Big|_{s=0}^{\infty}\right]\mathbf{F}^{\tra}(t), \label{genv}
\end{align}
where $\mathcal{G}$ is known as a \textit{response functional}, $\mathbf{F}$ is the deformation gradient and $\C=\mathbf{F}^{\tra}\mathbf{F}$ is the right Cauchy-Green deformation tensor. The stress will also, in general, be a function of space but for the sake of succinctness this argument is omitted.

In order to make progress, the form of the functional in \eqref{genv} clearly has to be specified. A fading memory hypothesis is generally assumed. This intuitive imposition simply states that more recent deformations or stresses are more important than those from the past. The most straightforward finite strain viscoelastic constitutive models are those of \textit{differential} type, assuming that the response functional is dependent on time derivatives of the right stretch tensor evaluated at the current time \cite{Lim-04, quintanilla2007importance, rashid2012mechanical}; however, more generality regarding the relaxation behaviour of the medium can be incorporated via integral forms. Single integral forms can be employed, which are essentially an extension of Boltzmann's superposition principle to finite deformations. Although such a principle clearly does not hold \textit{exactly} in a nonlinear setting, it can often provide a reasonable approximation. Furthermore, such an approach is often much more amenable to implementation than multiple integral forms \cite{Green1957, Fin-89, cheung1972nonlinear, darvish2001nonlinear}. Coleman and Noll \cite{coleman1961foundations} introduced \textit{finite linear viscoelasticity} where the response functional is linear in the Green strain. A model that has gained traction since its introduction, especially in recent times due to its flexibility, is the single integral Pipkin-Rogers model \cite{Pipkin-Rogers68}, which, if deformation is considered to begin at $t=0$, takes the form \cite{rajagopal2009response}
\begin{align}
\mathbf{T}(t) &=  \mathbf{F}(t)\left[\mathbf{Q}[\mathbf{C}(t),0] + \int_0^t \deriv{}{(t-s)}\mathbf{Q}[\mathbf{C}(s),t-s]\hspace{0.15cm} \text{d}s\right]\mathbf{F}^{\tra}(t), \label{PR}
\end{align}
with the first term being associated with the instantaneous elastic response and where $\mathbf{Q} = 2\partial\mathcal{W}/\partial\mathbf{C}$ for some potential $\mathcal{W}=\mathcal{W}(\mathbf{C}(s),t-s)$. This model has the advantage of allowing for strong nonlinearity and finite deformation. It also incorporates coupling between relaxation and strain, where necessary, although important decisions regarding the dependence of the potential $\mathcal{W}$ on the explicit time term $t-s$ must be made, motivated by experiments.

The theory of quasi-linear viscoelasticity (QLV), whose original form was proposed by Fung \cite{fung1972stress, Fun-81}, is a special case of \eqref{PR}. It incorporates finite strains and assumes that $\mathbf{Q}$ has a time/deformation separation so that
\begin{align}
\mathbf{T}(t) =\mathbf{F}(t)\left[\int^{t}_{-\infty}\mathbb{\tilde{G}}(t-s):\deriv{\PPi^{\textrm{e}}(\mathbf{C}(s))}{s}\text{d}s \right]\mathbf{F}^{\tra}(t), \label{1QLV}
\end{align}
where $\PPi^{\textrm{e}}=2\partial W/\partial\mathbf{C}$ is interpreted as the \textit{elastic} second Piola-Kirchhoff stress and $W$ is then the usual elastic strain energy function employed for finite elasticity problems. The notation $:$ indicates the double contraction between a fourth-order and a second-order tensor, such that $(\mathbb{A}:\textbf{B})_{ij}\!=\!A_{ijkl}B_{lk}$ in Cartesian coordinates. The term involving $\PPi^{\textrm{e}}$ forms an auxiliary measure of the strain in the medium. The tensor $\mathbb{\tilde{G}}$ is a time-dependent, reduced (non-dimensional) relaxation function tensor. In \cite{de2014nonlinear}, the \textit{isotropic} theory of QLV was revisited and the theory reformulated. The authors pointed out that a number of concerns raised recently as regards the efficacy of QLV were, in fact, unfounded.

QLV offers an attractive approach to modelling nonlinear viscoelastic materials that can be implemented rather straightforwardly in computations. A particularly attractive aspect of QLV is that relaxation functions can be determined from experiments in the linear viscoelastic regime. Although this restricts the constitutive form to modelling materials that do not exhibit strain dependent relaxation, it does immediately reduce the complexity of the model. Significant interest has focused on the case of transverse isotropy in the QLV context due to its importance in the application of modelling soft tissues. Small-strain QLV analyses have been conducted (see \cite{abramowitch2004improved}, for example), but the main focus for soft tissues has to be finite strains. Until now, the general trend in QLV theory has been to employ a \textit{scalar} relaxation function implementation, i.e.\ letting $\mathbb{\tilde{G}}(t-s)=G(t-s)\mathbb{I}$, where $G$ is a scalar function and $\mathbb{I}$ is the fourth-order identity tensor. This has also generally been the case in the isotropic case, as was pointed out in \cite{de2014nonlinear}. As key examples in the transversely isotropic (TI) scalar relaxation function QLV context, Huyghe \textit{et al.}\ \cite{huyghe1991constitutive} considered such an implementation for heart muscle tissue, Puso and Weiss \cite{puso1998finite} studied ligaments, Sahoo et \textit{al.}\ \cite{sahoo2014development} and Chatelin et \textit{al.}\ \cite{chatelin2013anisotropic} studied the brain, Motallebzadeh \textit{et al.}\ \cite{motallebzadeh2013non}, the eardrum and Jennesar \textit{et al.}\ \cite{jannesar2016transverse} focused on the spinal cord under tension. Vena \textit{et al.}\ implemented incompressible QLV with a scalar relaxation function but with separate relaxation contributions from fibres and matrix \cite{vena2006constituent}.
It is important to stress that there is no reason to expect that the relaxation response of complex viscoelastic materials should be the same in all modes of deformation in general. Indeed, even in an isotropic scenario, the hydrostatic and deviatoric relaxation response are almost always very different in their nature, not only in their relaxation spectra but even in their functional form \cite{tschoegl2012phenomenological}.

As pointed out by Weiss \textit{et al.}\ \cite{weiss2002ligament}, the \textit{in vivo} distribution of stress and strain in ligaments and tendons is highly inhomogeneous. This is also true of many other deformed soft tissues \textit{in-vivo}. Simulations are important for many reasons, but in particular for simulating surgery \cite{delingette2004soft}. An appropriate, fully three-dimensional constitutive model is therefore extremely important for accurate stress and deformation predictions. This is a fundamental motivation of the research carried out in the articles referred to above as well as of that presented here.

Work incorporating more than one relaxation function was carried out by Miller \textit{et al.}\ who employed Ogden-type polynomial expansions for $W$, in a QLV framework where relaxation terms accommodate distinct relaxation times depending upon the order of the term in the expansion \cite{miller1997constitutive, miller1999constitutive}. Since, however, these different orders are not associated with any specific physical deformations, they are incorporated purely to curve-fit to experiments. A large body of work that implicitly incorporates more than one relaxation function in anisotropic models is that associated with internal variable viscoelasticity theory, which was motivated by some of the earliest work on finite strain in viscoelastic isotropic solids \cite{valanis1971irreversible, lubliner1985model, simo1987fully}. This framework employs the uncoupled volumetric/deviatoric elasticity split dating back to Flory \cite{flory1961thermodynamic} and associates the time-dependent viscoelastic response to the deviatoric part only. The free energy function thus comprises a volumetric and isochoric elastic response, as well as a contribution due to configurational free energy associated with viscoelasticity. The decoupled stress then consists of equilibrium and non-equilibrium parts. The latter, which are described by evolving internal variables, dictate the viscoelastic response, and are governed by rate equations motivated by the linear theory \cite{valanis1971irreversible}. This theory was driven forward by Holzapfel and colleagues, who extended the work to more realistic strain energy functions, beyond the Gaussian network theory \cite{holzapfel1996large} and to anisotropic solids \cite{holzapfel2001viscoelastic, holzapfel2002structural}, always providing highly detailed analysis of finite element implementations. Pe\~{n}a \textit{et al.}\ developed this theory for TI materials in particular and applied it to ligaments and tendons \cite{pena2007anisotropic}. More recently, anisotropic viscoelasticity based on internal variable theory has been employed to model the eye \cite{Whi-18}, and, more generally, in modelling soft tissues \cite{Jer-18}. The internal variable approach has much in common with the QLV methodology, as was discussed in the isotropic case in \cite{de2014nonlinear}, and in particular, it exhibits time/deformation separation. In the anisotropic setting, however, since distinct relaxation functions do not appear explicitly, it is non-trivial to link them to specific physical modes of deformation. Very recent work studied the time/deformation separability assumption with reference to experiments on filled rubbers and a wide range of models \cite{Jri-18}. Interestingly, the internal variable model of Simo \cite{simo1987fully}, which is equivalent to QLV for isotropic, incompressible materials appears to fit data fairly well across a variety of experiments. Apparently, no such experimental data is yet available for anisotropic materials.

The detailed discussion above regarding the state-of-the-art on viscoelastic anisotropic theories motivates the work developed in this paper. A TI theory is proposed here, based on QLV, with a focus on the utility of the model, particularly in the incompressible regime, since this is a scenario of great practical importance. A tensor basis is employed for the relaxation tensor $\mathbb{\tilde{G}}$, which, in the incompressible limit, accommodates four relaxation functions. The proposed model is then used to predict the stress response in three common deformation modes: uni-axial extension along the fibres, and transverse and longitudinal shear. The results are presented for a specified strain energy function and specific relaxation functions in Section \ref{sec:res}. Moreover, we compare the proposed model with a standard isotropic QLV model to highlight the importance of including more than one relaxation function when modelling transversely isotropic soft tissues. We then use the TI QLV model to predict the Poynting effect in transversely isotropic materials. Our results give new insights on the role played by viscoelasticity in determining the Poynting effect for such materials. Conclusions are drawn in Section \ref{sec:concl}.

\section{Linear viscoelasticity} \label{sec:linvisc}
Anisotropic linear \textit{elastic} materials are characterised by their tensor of elastic moduli $\mathbb{C}$ with components $\mathbb{C}_{ijk\ell}$ in Cartesian coordinates. This tensor possesses the symmetries $\mathbb{C}_{ijk\ell}=\mathbb{C}_{jik\ell}=\mathbb{C}_{ij\ell k}=\mathbb{C}_{k\ell ij}$ and the tensor relates elastic stress to strain in the form $\si_{ij}^{\textrm{e}} = \mathbb{C}_{ijk\ell} \epsilon_{k\ell}$. In the case of TI materials, where the axis of anisotropy is in the direction of the unit vector $\mathbf{M}$, a classical form is the following (used extensively in the context of fibre-reinforced materials \cite{Spencer1984}):
\begin{align}
\boldsymbol{\si}^{\textrm{e}} &= \left(\la\tn{tr}\boldsymbol{\epsilon} + \alpha \epsilon_{\parallel}\right)\mathbf{I} + \left(\al\tn{tr}\boldsymbol{\epsilon}+\be\epsilon_{\parallel}\right)\mathbf{I}_{\parallel}+2\mu_T (\boldsymbol{\epsilon}-\boldsymbol{\epsilon}_{M}) + 2\mu_L\boldsymbol{\epsilon}_{M}, \label{TIHookea}
\end{align}
where $\mathbf{I}_{\parallel}=\mathbf{M}\otimes\mathbf{M}, \epsilon_{\parallel}=\mathbf{M}\cdot\boldsymbol{\epsilon}\mathbf{M}, \boldsymbol{\epsilon}_M=(\boldsymbol{\epsilon}(t)\mathbf{M})\otimes\mathbf{M} + \mathbf{M}\otimes(\boldsymbol{\epsilon}(t)\mathbf{M})$ and $\lambda$, $\alpha$, $\beta$ $\mu_T$ and $\mu_L$ are the elastic constants. A tensor basis can be employed in order to write $\mathbb{C}$ in a convenient form. This choice is non-unique, but for the form \eqref{TIHookea}, the modulus tensor can be written as
\begin{align}
\mathbb{C} &= \sum_{n=1}^6 j_n \mathbb{J}^n, \label{Spencerbasis}
\end{align}
where $j_1=\la$, $j_2=j_3=\alpha$, $j_4=\be$, $j_5=\mu_T$, $j_6=\mu_L$ and the basis tensors $\mathbb{J}^n, n=1,2,...,6$ are given in Appendix \ref{app:tens}. The choice of basis is motivated principally by the modes of deformation of interest and by which set of elastic moduli one wishes to work with.

In the incompressible limit, $\tn{tr}\boldsymbol{\epsilon}\rightarrow 0$ and $\lambda\rightarrow\infty$ such that \eqref{TIHookea} becomes
\begin{align}
\boldsymbol{\si}^{\textrm{e}} &= -p\mathbf{I} + \be\epsilon_{\parallel}\mathbf{I}_{\parallel}+2\mu_T (\boldsymbol{\epsilon}-\boldsymbol{\epsilon}_M) + 2\mu_L\boldsymbol{\epsilon}_M, \label{TIHookeincomp}
\end{align}
where we note that the term $\alpha\epsilon_{\parallel}$ in \eqref{TIHookea} can be incorporated into the Lagrange multiplier term $-p$ in \eqref{TIHookeincomp}. This means that there are only three independent elastic moduli for an incompressible, TI, linear elastic medium. Physically, this is explained by the fact that the restriction of zero volume change is not a purely hydrostatic condition (as is the case in an isotropic material) - the requirement means that the extension along the axis of anisotropy, $\mathbf{M}$ here, must also be constrained.

The constitutive form for anisotropic linear viscoelasticity theory for small strains, assuming fading memory and Boltzmann's superposition principle, takes the form \cite{christensen2012}
\begin{equation}\label{Fungforma}
\boldsymbol{\si}(t) = \int_{-\infty}^t \mathbb{R}(t-\tau):\deriv{\boldsymbol{\epsilon}}{\tau}\hspace{0.05cm} \text{d}\tau, 
\end{equation}
where the fourth-order relaxation tensor $\mathbb{R}$ has the symmetries $\mathbb{R}_{ijk\ell}(t)\!=\!\mathbb{R}_{jik\ell}(t)\!=\!\mathbb{R}_{ij\ell k}(t)$, but importantly, it does not, in general, possess the major symmetry $\mathbb{R}_{ijk\ell}(t)\!\neq \!\mathbb{R}_{k\ell ij}(t)$ \cite{shu2014anisotropic}. In the context of transverse isotropy, the tensor $\mathbb{R}(t)$ can be written in terms of TI tensor bases in the form
\begin{equation}\label{gJbasis}
\mathbb{R}(t) = \sum_{n=1}^6 R_n(t) \mathbb{J}^n, 
\end{equation}
where, with reference to \eqref{TIHookea}, $R_n(t)$ are time-dependent relaxation functions (with dimensions of stress), chosen such that
\begin{align}
R_1(0)&=\lambda, \quad R_2(0)=R_3(0)=\alpha,\quad R_4(0)=\beta,\quad R_5(0)=2\mu_T,\quad R_6(0)=2\mu_L,
\end{align}
and
\begin{align}
\lim_{t\rightarrow\infty}R_1(t)&=\lambda_\infty, \quad \lim_{t\rightarrow\infty}R_2(t)=\lim_{t\rightarrow\infty}R_3(t)=\alpha_\infty,\nonumber
\end{align}
\begin{align}
\lim_{t\rightarrow\infty}R_4(t)=\beta_\infty,\quad \lim_{t\rightarrow\infty}R_5(t)=2\mu_{T\infty},\quad \lim_{t\rightarrow\infty}R_6(t)=2\mu_{L\infty},
\end{align}
where $\lambda_\infty$, $\alpha_\infty$, $\beta_\infty$, $\mu_{T\infty}$ and $\mu_{L\infty}$ are the long-time moduli corresponding to $\lambda$, $\alpha$, $\beta$, $\mu_T$ and $\mu_L$, respectively. We note that these restrictions require $R_2(t)$ and $R_3(t)$ to be equal in the instantaneous and long-time limits; however, in general, they may relax at different rates and so cannot be assumed to be equal for all values of $t$. This further distinguishes the viscoelastic theory, which therefore has \textit{six} independent relaxation functions, from the elastic theory, which only has \textit{five} independent elastic constants.

A common choice is to let the $R_n(t)$ take the form of Prony series. A one-term Prony series for $R_1(t)$, for example, would be given by
\begin{equation}
 R_1(t)=\lambda_\infty+(\lambda-\lambda_\infty)e^{-t/\tau_1},
\end{equation}
where $\tau_1$ is the relaxation time associated with $\lambda$. The explicit form of \eqref{Fungforma} with \eqref{gJbasis} is
\begin{equation}\label{siLINTI}
\begin{split}
\boldsymbol{\si}(t) &= \int_{-\infty}^t \left(R_1(t-\tau)\deriv{}{\tau}\tn{tr}\boldsymbol{\epsilon}(\tau)
+ R_2(t-\tau)\deriv{}{\tau}\epsilon_{\parallel}(\tau)\right) \hspace{0.05cm}\text{d}\tau \hspace{0.1cm}\mathbf{I}  \\
& + \int_{-\infty}^t \left(R_3(t-\tau)\deriv{}{\tau}\tn{tr}\boldsymbol{\epsilon}(\tau) + R_4(t-\tau)\deriv{}{\tau}\epsilon_{\parallel}(\tau)\right) \hspace{0.05cm}\text{d}\tau \hspace{0.1cm}\mathbf{I}_{\parallel} 
 + \int_{-\infty}^t R_5(t-\tau)\deriv{}{\tau}(\boldsymbol{\epsilon}(\tau)-\boldsymbol{\epsilon}_{M}(\tau))\hspace{0.05cm}\text{d}\tau \\
 &+  \int_{-\infty}^t R_6(t-\tau)\deriv{}{\tau}\boldsymbol{\epsilon}_{M}(\tau)\hspace{0.05cm}\text{d}\tau, 
 \end{split}
\end{equation}
and in the incompressible limit this becomes
\begin{multline}
\boldsymbol{\si}(t)= -p(t)\mathbf{I} + \int_{-\infty}^t  R_4(t-\tau)\deriv{}{\tau}\epsilon_{\parallel}(\tau) \hspace{0.05cm}\text{d}\tau \hspace{0.1cm}\mathbf{I}_{\parallel} 
 + \int_{-\infty}^t R_5(t-\tau)\deriv{}{\tau}(\boldsymbol{\epsilon}(\tau)-\boldsymbol{\epsilon}_{M}(\tau))\hspace{0.05cm}\text{d}\tau +  \int_{-\infty}^t R_6(t-\tau)\deriv{}{\tau}\boldsymbol{\epsilon}_{M}(\tau)\hspace{0.05cm}\text{d}\tau. \label{siLINTIINCOMP}
\end{multline}
Next, with a view to the development of a modified quasi-linear theory of viscoelasticity for TI materials in the large deformation regime, i.e.\ in the form of \eqref{1QLV}, let us write the \textit{linear} TI viscoelastic constitutive equation in the form
\begin{align}
\boldsymbol{\si}(t) = \int_{-\infty}^t \mathbb{G}(t-\tau):\deriv{\boldsymbol{\si}^{\textrm{e}}}{\tau}\hspace{0.05cm} \text{d}\tau, \label{Fungform}
\end{align}
noting that we have now written $\boldsymbol{\si}^{\textrm{e}}$, the elastic (instantaneous) stress introduced in \eqref{TIHookea}, under the integral and therefore introduced the \textit{reduced} (non-dimensional) relaxation tensor $\mathbb{G}$. The choice of basis employed for $\mathbb{G}$ in \eqref{Fungform} is important since the time derivative of the instantaneous elastic stress is present under the integral and so in the elastic limit one should recover the pure elastic stress. To see this, we assume that deformation begins at $t=0$, and therefore integrate \eqref{Fungform} by parts to obtain
\begin{align}
\boldsymbol{\si}(t) = \mathbb{G}(0):\boldsymbol{\si}^{\textrm{e}}(t) +
\int_0^t \mathbb{G}'(t-\tau):\boldsymbol{\si}^{\textrm{e}}(\tau)\hspace{0.05cm} \text{d}\tau. \label{FungformB}
\end{align}
In order for us to recover the correct elastic limit at $t=0$ we must impose the condition $\mathbb{G}(0)=\mathbb{I}$, recalling that $\mathbb{I}$ is the fourth order identity tensor, with components $\mathbb{I}_{ijk\ell}=(\de_{ik}\de_{j\ell}+\de_{i\ell}\de_{jk})/2$ in Cartesian coordinates. Assuming a tensor basis decomposition for $\mathbb{G}$ of the form
\begin{align}
\mathbb{G}(t) &= \sum_{n=1}^6 G_n(t)\mathbb{K}^n, \label{GTI}
\end{align}
for some new tensor basis $\mathbb{K}^n, n=1,2,...,6$, with $G_n(0)=1$ for all $n$, this means that we must have
\begin{align}
\sum_{n=1}^6 \mathbb{K}^n &= \mathbb{I}. \label{ID}
\end{align}
The basis $\mathbb{J}^n$ introduced above does \textit{not} have this property, but a basis for transverse isotropy that \textit{does} is given in \eqref{basisTI1}-\eqref{basisTI4} of Appendix \ref{app:tens}. Using \eqref{GTI} and \eqref{TIHookea} in \eqref{Fungform}, and equating the resulting expression to \eqref{Fungforma}, yields the connections between $R_n$ and $G_n$, which are given explicitly  in \eqref{Gg3}. Finally, the basis $\mathbb{K}^n$ decomposes a second-order tensor as follows:
\begin{align}
 \boldsymbol{\si}^{\textrm{e}} &= \sum_{n=1}^6 \mathbb{K}^n:\boldsymbol{\si}^{\textrm{e}} =  \boldsymbol{\si}^{\textrm{e}}_1 + \boldsymbol{\si}^{\textrm{e}}_2 + \boldsymbol{\si}^{\textrm{e}}_3 + \boldsymbol{\si}^{\textrm{e}}_4 + \boldsymbol{\si}^{\textrm{e}}_5 + \boldsymbol{\si}^{\textrm{e}}_6,
\end{align}
where the explicit forms of the terms $\boldsymbol{\si}^{\textrm{e}}_n$ are stated in \eqref{sigsplit1}-\eqref{SIGS}. The expression \eqref{Fungform} is now employed as the basis for a modified QLV theory for nonlinear materials subject to finite deformations. In this regime, appropriate stress measures that account for finite strains have to be used, as shall now be discussed.

\section{Modified quasi-linear viscoelasticity} \label{sec:QLV}
The viscoelasticity theory described above is now extended in order to deal with materials that are subject to finite deformation and whose constitutive response is nonlinear. In particular, a modified QLV theory is developed where relaxation is independent of deformation. Initially, the theory will be developed for general, compressible materials before the incompressible limit is taken. This yields a relatively compact constitutive model for incompressible TI viscoelastic materials that is suitable for use in computational models and for further development to model more complex materials.

\subsection{General compressible form}
We begin by defining $\XX=\sum^{3}_{i=1}X_i\EE_i$ and $\xx(t)=\sum^{3}_{j=1}x_j(t)\ee_j$ as the position vectors that identify a point of the body in the initial configuration (at $t=0$) and current configurations of the body, $\bb_0$ and $\bb(t)$, respectively. The deformation gradient $\ff(t)$ is defined as
\begin{equation}\label{def}
\ff(t)=\Grad\xx(t)=\dfrac{\partial\xx(t)}{\partial \XX}, \quad \text{with}\quad \ff(t\leq 0)=\textbf{I}\quad\text{and}\quad J=\det \ff.
\end{equation}
The left (right) Cauchy strain tensor is defined as $\B=\ff\ff^{\tra}$ ($\C=\ff^{\tra}\ff$) and its three isotropic invariants are given by
\begin{equation}\label{isoI}
I_1=\tr \B\quad I_2=\dfrac{1}{2}(I_1^2-\tr \B^2)\quad I_3=\det \B.
\end{equation}
Let $\M$ and $\m=\ff\M$ be the vectors along the principal axis of anisotropy of the TI material in question in the undeformed and deformed configurations, respectively. The anisotropy could be associated with, for example, the direction of the axis of aligned fibres in a medium, the direction orthogonal to parallel layers, or something more complex.  Transverse isotropy requires an additional two anisotropic invariants
\begin{equation}\label{anisoI}
I_4=\m\cdot\m \quad\text{and}\quad I_5=\m\cdot(\B\m).
\end{equation}
By assuming the existence of a strain energy function $W(I_i)$ with $i=\{1,\dots,5\}$, the elastic Cauchy stress $\textbf{T}^{\textrm{e}}=J^{-1}\ff\,\sum^{5}_{i=1}W_i\dfrac{\partial\,I_i}{\partial\,\ff}$ can be written in the general form \cite{ogden2007incremental}
\begin{equation}\label{nonlinCS}
\textbf{T}^{\textrm{e}}=2J^{-1}\Big(I_3W_3\textbf{I}+W_1\B-W_2\B^{-1}+W_4\m\otimes\m+W_5(\m\otimes\B\m+\B\m\otimes\m)\Big),
\end{equation}
where the subscript $i$ denotes differentiation with respect to the $i$th strain invariant. The notation $\textbf{T}^{\el}$ distinguishes this nonlinear form from its linear counterpart \eqref{TIHookea}, which we call $\CS^{\textrm{e}}$.

QLV requires a constitutive model of a form similar to \eqref{Fungform}; however, the theory cannot be directly formulated in terms of the Cauchy stress since objectivity must be ensured \cite{de2014nonlinear}. Instead, \eqref{Fungform} can be written with respect to the second Piola-Kirchhoff stress tensor $\PPi(t)$ and its elastic counterpart $\PPi^{\el}\!=\!J\textbf{F}^{-1}\textbf{T}^{\el}\textbf{F}^{-\text{T}}$. Upon integrating by parts, the constitutive equation for QLV (see equation \eqref{1QLV}) can be written as
\begin{equation}\label{QLVEPPiold}
\PPi(t)=\mathbb{\tilde{G}}(0):\left(J(t)\textbf{F}^{-1}(t)\textbf{T}^{\el}(t)\textbf{F}^{-\tra}(\tau)\right)
+\int^{t}_0\mathbb{\tilde{G}}'(t-\tau):\left(J(\tau)\textbf{F}^{-1}(\tau)\textbf{T}^{\el}(\tau)\textbf{F}^{-\tra}(\tau)\right)\text{d}\tau,
\end{equation}
where $\mathbb{\tilde{G}}$ is a fourth-order reduced relaxation function tensor. If we were to split $\mathbb{\tilde{G}}$ in terms of fundamental bases, then, in the isotropic case, this would correspond to splitting the elastic second Piola-Kirchhoff stress tensor into its hydrostatic and deviatoric parts, which do not have a clear physical interpretation. Therefore, instead of this approach, we choose to apply the split to the Cauchy stress in the following manner:
\begin{equation}\label{QLVEPPi0}
\PPi(t)=J(t)\textbf{F}^{-1}(t)\left(\mathbb{G}(0):\textbf{T}^{\el}(t)\right)\textbf{F}^{-\tra}(t)
+\int^t_0 J(\tau)\textbf{F}^{-1}(\tau)\left(\mathbb{G}'(t-\tau):\textbf{T}^{\el}(\tau)\right)\textbf{F}^{-\tra}(\tau)\text{d}\tau,
\end{equation}
which, by using the TI bases introduced in \eqref{basisTI1}-\eqref{basisTI4} amounts to writing
\begin{equation}\label{QLVEPPi}
\PPi(t)=\sum^{6}_{n=1}G_n(0)\PPi^{\el}_n(t)+\,\int^{t}_{0}\sum^{6}_{n=1}G_n'(t-\tau)\PPi^{\textrm{e}}_n(\tau)\text{d}\tau,
\end{equation}
where $G_n(t)$ are the components of the relaxation function $\mathbb{G}$ introduced in Section \ref{sec:linvisc}, such that $G_n(0)=1$. This modified version of the QLV theory still preserves the property of the relaxation functions being independent of the deformation; however, the bases introduced in \eqref{basisTI1}-\eqref{basisTI4} do depend on the deformation through the vector $\textbf{m}\!=\!\ff\textbf{M}$.
The terms $\PPi^{\textrm{e}}_n$ are given by
\begin{equation}\label{splitPPi}
\PPi^{\textrm{e}}_n=J\ff^{-1}(\mathbb{K}_n:\textbf{T}^{\el})\ff^{-\!\tra}=J\ff^{-1}\textbf{T}^{\textrm{e}}_n\ff^{-\!\tra},\qquad \text{with}\qquad n=\{1,\ldots,6\},
\end{equation}
where $\textbf{T}^{\textrm{e}}_n$ are given by equations \eqref{sigsplit1}-\eqref{sigsplit3}, but with the components $\sigma^{\textrm{e}}_{ij}$ replaced by $T^{\textrm{e}}_{ij}$. We note that equation \eqref{QLVEPPi0} cannot be written in the form of equation \eqref{QLVEPPiold} for any choice of $\mathbb{\tilde{G}}$ and, therefore, this constitutive expression is \textit{not} the classical QLV approach; however, due to the similarities between the forms, we use the term \textit{modified QLV} to describe this expression. It should be noted that this comment also applies to the compressible isotropic theory developed in \cite{de2014nonlinear}.

The components $G_n(t)$ are related to the functions $R_n(t)$ (which are associated with the natural split of linear TI viscoelasticity in \eqref{siLINTI}) via equation \eqref{Gg3}. Note that the relaxation functions $G_n(t)$ are independent of deformation, which is a fundamental and important assumption of both QLV and modified QLV and it means that the constitutive equation is restricted to materials for which this is a good approximation. It does mean, however, that such functions can be measured directly by small-strain tests on the medium in question, which is an attractive aspect of the theory, as we shall discuss later.

The connection \eqref{Gg3} can now be employed to replace $G_n$ with $R_n$ in \eqref{QLVEPPi} and the Cauchy stress can then be written as:
\begin{equation}\label{QLVCauchy}
\textbf{T}(t)=J^{-1}(t)\ff(t)\sum^{6}_{n=1}R_n(0)\boldsymbol{P}_{n}^{\el}(t)\ff^{\tra}(t)
+J^{-1}\ff(t)\Big(\int^{t}_{0}\sum^{6}_{n=1}R_n'(t-\tau)\boldsymbol{P}^{\el}_n(\tau)\text{d}\tau\Big)\ff^{\tra}(t),
\end{equation}
where the $\boldsymbol{P}^{\el}_n, n=1,2,...,6$ are given in \eqref{bigSigma}. Equation \eqref{QLVCauchy} can be used to model compressible materials. In the next section, we shall consider the incompressible limit, since this is of great utility in a number of important applications, including polymer composites and soft tissues.

\subsection{The incompressible limit} \label{incomplimit}
We now have a model that can accommodate fully compressible TI behaviour in the large deformation regime. In order to illustrate the applicability of the model, let us consider the important case of incompressibility. The incompressible limit is recovered by setting $J\rightarrow 1$ and $\lambda\rightarrow\infty$. The elastic constitutive equation in \eqref{nonlinCS} in this limit is
\begin{equation}\label{CSincEL}
\mathbf{T}^{\textrm{e}}(t) = -p^{\el}(t)\mathbf{I} +2W_1\B-2W_2\B^{-1}+2W_4\m\otimes\m+2W_5(\m\otimes\B\m+\B\m\otimes\m).
\end{equation}
Moreover, from equations \eqref{app:A1}-\eqref{app:A2},
\begin{equation}\label{Bt}
\lim_{\lambda\rightarrow\infty}A=\lim_{\lambda\rightarrow\infty}C
=\lim_{\lambda\rightarrow\infty}D=0\qquad \text{and}\qquad \mathcal{B}=\lim_{\lambda\rightarrow\infty}B=\dfrac{1}{\beta+4\mu_L-\mu_T}=\dfrac{1}{E_L},
\end{equation}
where $E_L$ is the longitudinal Young modulus. For details on how to derive the last equality in \eqref{Bt} we refer to \cite{lubarda2008elastic} and the references therein. The incompressible limit of equation \eqref{QLVCauchy} then becomes
\begin{equation}\label{VEsigmaINC}
\begin{split}
\textbf{T}(t)&=-p(t)\textbf{I}+\tilde{T}^{\el}(t)\textbf{m}\otimes\textbf{m}+\textbf{T}^{\el}_5(t)+\textbf{T}^{\el}_6(t)
+\lim_{J\rightarrow1,\,\lambda\rightarrow\infty}\left(J^{-1}\ff(t)\Big(\int^{t}_{0}R_1'(t-\tau)\boldsymbol{P}^{\el}_1(\tau)\text{d}\tau\Big)\ff^{\tra}(t)\right)\\
&+J^{-1}\ff(t)\Big(\int^{t}_{0}R_2'(t-\tau)\,\boldsymbol{P}^{\el}_{2}(\tau)\text{d}\tau\Big)\ff^{\tra}(t)
+\ff(t)\Big(\int^{t}_{0}\dfrac{\mathcal{R}'(t-\tau)}{E_L}\,\PPi^{\el}_{\Lo}(\tau)\text{d}\tau\Big)\ff^{\tra}(t)\\
&+\ff(t)\Big(\int^{t}_{0}\dfrac{R_5'(t-\tau)}{2\mu_T}(\PPi^{\el}_{\T}(\tau)-\PPi_{\cc}^{\el}(\tau))\text{d}\tau\Big)\ff^{\tra}(t)
+\ff(t)\Big(\int^{t}_{0}\dfrac{R_6'(t-\tau)}{2\mu_L}\PPi^{\el}_{\A}(\tau)\text{d}\tau\Big)\ff^{\tra}(t),
\end{split}
\end{equation}
where the Lagrange multiplier $p(t)$ is given by:
\begin{equation}\label{p}
\begin{split}
-p(t)\textbf{I}&=\lim_{J\rightarrow1,\,\lambda \rightarrow\infty}\Big(\,R_1(0)\left(A\tilde{T}^{\el}(t)-C\bar{T}^{\el}(t)\right)+\,R_2(0)\left(B\tilde{T}^{\el}(t)-D\bar{T}^{\el}(t)\right)\\
&+R_5(0)\left(\dfrac{A\tilde{T}^{\el}(t)-C\bar{T}^{\el}(t)}{2}-\dfrac{B\tilde{T}^{\el}(t)-D\bar{T}^{\el}(t)}{2}\right)\Big)\textbf{I}.
\end{split}
\end{equation}
The first integral in \eqref{VEsigmaINC} vanishes if we assume that in the incompressible limit the time-dependence of this relaxation function becomes instantaneous so that $R_1(t)=\lambda$ for all $t$, and therefore $R_1'(t)=0$. Moreover, in \eqref{VEsigmaINC}, the relaxation function $\mathcal{R}(t)$ and the following terms have been introduced:
\begin{equation} \label{TERMS}
\begin{split}
\mathcal{R}(t) &= R_4(t)-\nicefrac{1}{2}\,R_5(t)+2R_6(t),\\
\PPi_{\T}^{\el} &= \ff^{-1}\TT^{\el}_5\ff^{-\!\tra}\\
\PPi_{\cc}^{\el} &= \ff^{-1}\left(\dfrac{\mu_T}{E_L}\tilde{T}^{\el}\textbf{I}\right)\ff^{-\!\tra},
\end{split}
\hspace{3em}
\begin{split}
\PPi_{\Lo}^{\el} &= \tilde{T}^{\el}\ff^{-1}\mathbf{m}\otimes\mathbf{m}\ff^{-\!\tra}, \\
\PPi_{\A}^{\el} &= \ff^{-1}\TT^{\el}_6\ff^{-\!\tra},
\end{split}
\end{equation}
where, as we shall show in the next section, $\mathcal{R}(t)$ is associated with relaxation in the direction of the axis of anisotropy $\mathbf{m}$, where $E_L$ is the axial Young's modulus. The subscripts $\T$ and $\A$ on $\PPi^{\textrm{e}}$ are associated with in-plane (transverse) shear in the plane of isotropy and anti-plane (longitudinal) shear, respectively. We note that $R_4(t)$, $R_5(t)$ and $R_6(t)$ all appear in \eqref{siLINTIINCOMP}, but $R_2(t)$ does not, therefore it is not possible to determine $R_2(t)$ from the \textit{incompressible} linear theory (although it can be determined in the compressible case). For simplicity, therefore, in the following sections, we shall make the constitutive assumption that $R_2(t)=0$ for all $t$, which restricts our attention to those incompressible TI materials for which this assumption is reasonable. The constitutive equation we use, therefore, is
\begin{equation}\label{VEsigmaINC2}
\begin{split}
\textbf{T}(t)&=-p(t)\textbf{I}+\tilde{T}^{\el}(t)\textbf{m}\otimes\textbf{m}+\textbf{T}^{\el}_5(t)+\textbf{T}^{\el}_6(t)
+\ff(t)\Big(\int^{t}_{0}\dfrac{\mathcal{R}'(t-\tau)}{E_L}\,\PPi^{\el}_{\Lo}(\tau)\text{d}\tau\Big)\ff^{\tra}(t)\\
&+\ff(t)\Big(\int^{t}_{0}\dfrac{R_5'(t-\tau)}{2\mu_T}(\PPi^{\el}_{\T}(\tau)-\PPi_{\cc}^{\el}(\tau))\text{d}\tau\Big)\ff^{\tra}(t)
+\ff(t)\Big(\int^{t}_{0}\dfrac{R_6'(t-\tau)}{2\mu_L}\PPi^{\el}_{\A}(\tau)\text{d}\tau\Big)\ff^{\tra}(t).
\end{split}
\end{equation}
Under the assumptions mentioned above, the stress can be written in terms of three relaxation functions: $\mathcal{R}(t), R_5(t)$ and $R_6(t)$. These can be determined independently via three linear viscoelastic tests associated with uniaxial loading, in-plane shear and anti-plane shear, respectively. This is an appeal of the model in the sense that the viscoelastic behaviour can be fully characterised by experiments in the linear viscoelastic regime (in the fully compressible case, three more experiments would be required to determine $R_1(t)$, $R_2(t)$ and $R_3(t)$). Of course, in reality, not all materials respond viscoelastically in this manner and the relaxation functions can depend on strain amplitude, in principle. For now, we accept that the model cannot incorporate this effect but emphasise that not only is it capable of incorporating large deformations, but that it can also accommodate distinct relaxation behaviours associated with anisotropy, as opposed to the vast majority of existing models. We note further that most of these existing models also do not incorporate strain-dependent relaxation.

\subsection{Linkage of relaxation functions to small-strain deformation modes} \label{EXPTS}

In this section we briefly summarise how, for an incompressible material, the three relaxation functions $\mathcal{R}(t),R_5(t),R_6(t)$ can be associated with three independent small-strain regime tests: a simple extension in the direction of the axis of anisotropy and two shear deformations: in-plane and anti-plane shear. A range of small strain time-dependent experiments then permit the experimental determination of these relaxation functions and their associated relaxation spectra.

Take the $\EE_3$-axis to be the axis of anisotropy. We first consider a scenario where a sample of TI material is stretched along this axis. The infinitesimal strain tensor is then $\boldsymbol{\epsilon}=\epsilon_{33}\ee_3\otimes\ee_3$, where $\epsilon_{33}$ is the strain component along the $\ee_3$ axis. The associated stress response from \eqref{siLINTIINCOMP} becomes:
\begin{equation}\label{linext}
\sigma_{33}(t)=\int_{-\infty}^t\mathcal{R}(t-\tau)\dfrac{\partial \,\epsilon_{33}(\tau)}{\partial\tau}\text{d}\tau.
\end{equation}
We note that equation \eqref{linext} depends on three relaxation functions (since $\mathcal{R}(t)=R_4(t)-\nicefrac{1}{2}\,R_5(t)+2R_6(t)$); therefore, from this test we are only able to obtain information about the composite relaxation function $\mathcal{R}(t-\tau)$, which is associated with uniaxial deformation. To explicitly determine all three relaxation functions, two further tests are required. An in-plane shear test can be carried out in the plane of isotropy. This deformation is associated with a strain tensor of the form $\boldsymbol{\epsilon}=\epsilon_{12}(\ee_1\otimes\ee_2+\ee_2\otimes\ee_1)$, where $\epsilon_{12}$ is the amount of shear in the isotropic plane. The resulting stress response is
\begin{equation}\label{linTshear}
\sigma_{12}(t)=\int_{-\infty}^tR_5(t-\tau)\dfrac{\partial\, \epsilon_{12}(\tau)}{\partial\tau}\text{d}\tau;
\end{equation}
therefore, from an in-plane shear test we can extract the parameters appearing in the relaxation function $R_5(t)$. The third and last test required is a shear deformation in one of the two planes containing the axis of anisotropy (either $\ee_1$-$\ee_3$ or $\ee_2$-$\ee_3$). Upon choosing the plane $\ee_1$-$\ee_3$, the strain tensor can be written as
$\boldsymbol{\epsilon}=\epsilon_{13}(\ee_1\otimes\ee_3+\ee_3\otimes\ee_1)$, where $\epsilon_{13}$ is the amount of shear in the $\ee_1$-$\ee_3$ plane. The shear stress response, then, is given by
\begin{equation}\label{linLshear}
\sigma_{13}(t)=\int_{-\infty}^tR_6(t-\tau)\dfrac{\partial\, \epsilon_{13}(\tau)}{\partial\tau}\text{d}\tau,
\end{equation}
which allows us to determine the parameters appearing in $R_6(t)$. Once $\mathcal{R}(t)$, $R_5(t)$ and $R_6(t)$ are known, they can be used to calculate $R_4(t)$ from the first equation of \eqref{TERMS}. As mentioned above, to use the compressible theory, three additional, similar experiments would need to be carried out in order to determine the relaxation functions $R_1(t)$, $R_2(t)$ and $R_3(t)$.

\section{Deformation}\label{sec:def}
Let us now consider specific deformations of the medium in question. As above, let us take the axis of anisotropy to be $\mathbf{M}=\mathbf{E}_3$, so that for a fibre-reinforced composite, for example, the fibres are all aligned along the $\EE_3$-direction in the undeformed configuration. In all cases, it is assumed  that the deformations are slow enough that inertial terms can be neglected and, therefore, since the deformations considered are homogeneous, they automatically satisfy the equations of motion. All deformations begin at $t=0$ and we use the notation $X_1, X_2, X_3$ and $x_1, x_2, x_3$ for Cartesian coordinates in the undeformed and deformed configurations, respectively. In Section \ref{sec:res}, we shall use the expressions derived in this section to plot relaxation curves for given strain energy functions.

\subsection{Uniaxial deformation}\label{sec:exte}
First, consider the following deformation, which is associated with \textit{uniaxial deformation} under tension with no lateral traction. Assume that the deformation begins at $t=0$ and for $t\geq 0$, we have
\begin{align}\label{exte}
x_1(t)&=1/\sqrt{\Lambda(t)}X_1, &
x_2(t)&=1/\sqrt{\Lambda(t)}X_2 &
\textnormal{and}& & x_3(t)&=\Lambda(t)X_3.
\end{align}
This corresponds to a simple extension, with stretch $\Lambda(t)$, in the direction of the axis of anisotropy.  If anisotropy is induced by fibres aligned along the $\EE_3$ direction, equation \eqref{exte} represents a simple extension in the direction of the fibres. Assuming that this deformation has been generated by a non-zero axial stress $T_{33}$ with the lateral surfaces being free of traction, the stress state is assumed to take the form
\begin{align}\label{bcexte}
T_{33} &= T(t), & T_{11}=T_{22} &=0 & \textnormal{and} & & T_{ij}=0 \hspace{0.2cm}(i\neq j).
\end{align}
The deformation gradient associated with equation \eqref{exte} is diagonal:
\begin{equation}\label{Fexte}
\ff(t)=\diag(1/\sqrt{\Lambda(t)},1/\sqrt{\Lambda(t)},\Lambda(t)),
\end{equation}
for $t>0$, which gives rise to the following left Cauchy-Green strain tensor
\begin{equation}\label{Bexte}
\B(t)=\diag(1/\Lambda(t),1/\Lambda(t),\Lambda^2(t)).
\end{equation}
From equation \eqref{VEsigmaINC2}, the stress $T(t)$ is given by
\begin{equation}\label{CS33}
\begin{split}
T(t)&=-p(t)\\
&+\Lambda^2 (t)\left(\tilde{T}^{\el}(t)+\int^{t}_0\dfrac{\mathcal{R}'(t-\tau)}{E_L} \Pi^{\el}_{\Lo 33}(\tau)\text{d}\tau-\int^{t}_0\dfrac{R_5'(t-\tau)}{2\mu_T} \Pi^{\el}_{\cc 33}(\tau)\text{d}\tau\right).
\end{split}
\end{equation}
The Lagrange multiplier $p(t)$ can be calculated from the second equation of \eqref{bcexte}:
\begin{equation}\label{pexte}
p(t)= -\dfrac{1}{\Lambda(t)}\int^t_0\dfrac{R_5'(t-\tau)}{2 \mu_T} \Pi^{\el}_{\cc 11}(\tau) \text{d}\tau.
\end{equation}
Furthermore,
\begin{align}\label{PiExte}
\Pi_{\Lo 33}^{\textrm{e}}(\tau) &=\tilde{T}^{\el}(\tau),
&\Pi_{\cc 33}^{\textrm{e}}(\tau) &= \dfrac{\mu_T}{ E_L}\dfrac{1}{\Lambda^2(\tau)}\tilde{T}^{\el}(\tau),
&\Pi^{\el}_{\cc 11}(\tau)=\dfrac{\mu_T}{E_L}\,\tilde{T}^{\el}(\tau)\Lambda(\tau),
\end{align}
where the term $\tilde{T}^{\el}$ is defined analogously to $\tilde{\si}^{\el}$ in equation \eqref{SIGS1} and can be calculated by combining equations \eqref{CSincEL} and \eqref{Bexte}. Upon substituting equations \eqref{pexte} and \eqref{PiExte} into equation \eqref{CS33}, we obtain
\begin{equation}\label{finalstressexte0}
T_{33}(t)=\Lambda^2 (t)\Big(\tilde{T}^{\el}(t)+\int^{t}_0\dfrac{\mathcal{R}'(t-\tau)}{E_L} \tilde{T}^{\el}(\tau)\text{d}\tau\Big)
+\int_0^t \dfrac{R_5'(t-\tau)}{2\mu_T}\dfrac{\mu_T}{E_L}\left(\dfrac{\Lambda(\tau)}{\Lambda(t)}-\dfrac{\Lambda^2(t)}{\Lambda^2(\tau)}\right)\tilde{T}^{\el}(\tau)\text{d}\tau.
\end{equation}
We note that equation \eqref{finalstressexte0} depends only on $R_4(t)$, $R_5(t)$ and $R_6(t)$, as was the case for its linear counterpart in \eqref{linext} and, in the small-strain limit, \eqref{finalstressexte0} becomes identical to \eqref{linext}.

\subsection{In-plane (transverse) shear}\label{sec:ts}
\begin{figure}[b!]
\centering
\subfigure[Undeformed block]{
	\centering
	\includegraphics[scale=0.6]{shear}}
\qquad
\subfigure[Transverse shear]{
	\centering
	\raisebox{16pt}{
	\includegraphics[scale=0.6]{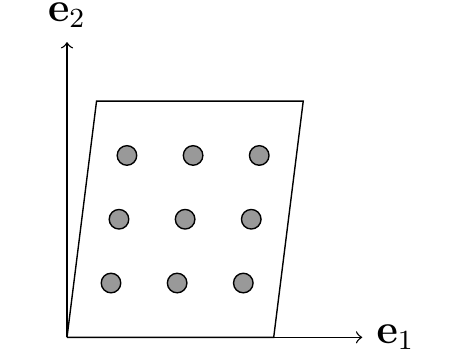}}}
\caption{A block of a TI material with fibres pointing in the direction of the $\EE_3$-axis under the simple transverse shear deformation in \eqref{ts}.}\label{fig:Tsh}
\end{figure}

Let us now consider a homogeneous simple shear deformation in the plane of isotropy $\ee_1$-$\ee_2$, as sketched in Figure \ref{fig:Tsh}.
This type of shear is often called \textit{transverse shear} \cite{MURPHY201390,destrade2015dominant}. The deformation is written as
\begin{align}\label{ts}
x_1(t)&=X_1+\kappa(t)X_2,  & x_2(t)&=X_2, & x_3(t) &= X_3,
\end{align}
where $\kappa(t)$ is the amount of shear. In this case, the deformation gradient $\ff(t)$ and the left Cauchy-Green strain tensor $\B(t)$ are given by
\begin{equation}\label{FBts}
\ff(t)=\left(
\begin{array}{ccc}
1 &\kappa(t) &0\\
0 &1 &0\\
0 &0 &1
\end{array}\right)
\qquad \text{and}\qquad
\B(t)=\left(
\begin{array}{ccc}
1+\kappa^2(t) &\kappa(t) &0\\
\kappa(t) &1 &0\\
0 &0 &1
\end{array}\right).
\end{equation}
A traction-free boundary condition is imposed on the surface with normal $\mathbf{N}=(0,0,1)$, which implies that
\begin{equation}\label{bcts}
T_{13}=T_{23}=T_{33}=0, \qquad \qquad \forall t.
\end{equation}
The non-zero components of the Cauchy stress tensor from equation \eqref{QLVCauchy} are then
\begin{equation}\label{sigmats}
\begin{split}
T_{11}(t)&=-p(t)+\dfrac{T^{\el}_{11}(t) -  T^{\el}_{22}(t) }{2}+\int^t_0\dfrac{R_5'(t-\tau)}{2\mu_T}\Big(\Pi^{\el}_{\T 11}(\tau)-\Pi^{\el}_{\cc 11}(\tau)\\
&+ 2 \kappa(t) \left(\Pi^{\el}_{\T 12}(\tau)-\Pi^{\el}_{\cc 12}(\tau)\right)
 + \kappa^2(t) \left(\Pi^{\el}_{\T 22}(\tau)-\Pi^{\el}_{\cc 22}(\tau)\Big)  \right)\text{d}\tau,\\
T_{22}(t)&=-p(t)+\dfrac{T^{\el}_{22}(t) - T^{\el}_{11}(t) }{2}
+\int^t_0\dfrac{R_5'(t-\tau)}{2\mu_T}\left(\Pi^{\el}_{\T 22}(\tau)-\Pi^{\el}_{\cc 22}(\tau)\right)\text{d}\tau,\\
T_{12}(t)&=T^{\el}_{12}(t)
+\int^t_0\dfrac{R_5'(t-\tau)}{2\mu_T}\Big(\Pi^{\el}_{\T 12}(\tau)-\Pi^{\el}_{\cc 12}(\tau)+\kappa(t)\left(\Pi^{\el}_{\T 22}(\tau)-\Pi^{\el}_{\cc 22}(\tau)\right)\Big)\text{d}\tau,
\end{split}
\end{equation}
where the Lagrange multiplier $p(t)$ can be calculated from the last equation of \eqref{bcts} as follows:
\begin{equation}\label{Lagts}
\begin{split}
 p(t)= \tilde{T}^{\el}(t)+\int^t_0\dfrac{\mathcal{R}'(t-\tau)}{E_L} \Pi^{\el}_{\Lo33}(\tau) \text{d}\tau-\int^t_0\dfrac{R_5'(t-\tau)}{2 \mu_T} \Pi^{\el}_{\cc 33}(\tau) \text{d}\tau.
\end{split}
\end{equation}
Moreover, we have
\begin{equation}\label{Sigmats}
\begin{split}
\Pi^{\el}_{\cc 11}&=\nicefrac{\mu_T}{E_L}(1 + \kappa^2)  \tilde{T}^{\el},
\hspace{1.5em}
\Pi^{\el}_{\cc 22}=\nicefrac{\mu_T}{E_L}\tilde{T}^{\el},\hspace{1.5em}
\Pi^{\el}_{\cc 12}=-\nicefrac{\mu_T}{E_L}\kappa \tilde{T}^{\el},\hspace{1.5em}
\Pi^{\el}_{\cc33}=\nicefrac{\mu_T}{E_L}\tilde{T}^{\el},\\
\Pi^{\el}_{\T11}&=\dfrac{T^{\el}_{22} -T^{\el}_{11}}{2}(\kappa^2-1)-2\kappa T^{\el}_{12},
\hspace{1.3em}
\Pi^{\el}_{\T22}=\dfrac{T^{\el}_{22} -T^{\el}_{11}}{2},\hspace{1.3em}
\Pi^{\el}_{\T12}=T^{\el}_{12} - \kappa \dfrac{T^{\el}_{22} -T^{\el}_{11}}{2},\hspace{1.3em}
\Pi^{\el}_{\Lo33}=\tilde{T}^{\el}.
\end{split}
\end{equation}
Finally, substituting equations \eqref{Lagts} and \eqref{Sigmats} into equation \eqref{sigmats}, we obtain the stress components $T_{11},T_{22}$ and $T_{12}$:
\begin{equation}\label{sigmatsfinaladim}
\begin{split}
T_{11}(t)&=T^{\el}_{11}(t)-T^{\el}_{33}(t)-\int^t_0\dfrac{\mathcal{R}'(t-\tau)}{E_L}\tilde{T}^{\el}(\tau)\text{d}\tau
-\int^t_0\dfrac{R_5'(t-\tau)}{2\mu_T}\dfrac{T^{\el}_{22}(\tau)-T^{\el}_{11}(\tau)}{2}\text{d}\tau\\
&+\int^t_0\dfrac{R_5'(t-\tau)}{2\mu_T}\left(\dfrac{T^{\el}_{22}(\tau)-T^{\el}_{11}(\tau)}{2}-\dfrac{\mu_T}{E_L}\tilde{T}^{\el}(\tau)\right) \left(\kappa(t) - \kappa(\tau)\right)^2\text{d}\tau 
+2 \int^t_0\dfrac{R_5'(t-\tau)}{2\mu_T} \left(\kappa(t)-\kappa(\tau)\right) T^{\el}_{12}(\tau) \text{d}\tau,\\
T_{22}(t)&=T_{22}^{\el}(t)-T_{33}^{\el}(t)-\int^t_0\dfrac{\mathcal{R}'(t-\tau)}{E_L}\tilde{T}^{\el}(\tau)\text{d}\tau
+\int^t_0\dfrac{R_5'(t-\tau)}{2\mu_T}\left(\dfrac{T_{22}^{\el}(\tau)-T_{11}^{\el}(\tau)}{2}\right)\text{d}\tau,\\
T_{12}(t)&=T_{12}^{\el}(t)+\int^t_0\dfrac{R_5'(t-\tau)}{2\mu_T} T^{\el}_{12}(\tau) \text{d}\tau
+\int^t_0\dfrac{R_5'(t-\tau)}{2\mu_T} \left(\dfrac{T^{\el}_{22}(\tau) - T^{\el}_{11}(\tau)}{2}-\dfrac{\mu_T}{E_L}\tilde{T}^{\el}(\tau) \right)\left(\kappa(t)-\kappa(\tau)\right)\text{d}\tau.
\end{split}
\end{equation}
The corresponding elastic stresses $T_{11}^{\el},T_{22}^{\el},T_{12}^{\el}$ and $\tilde{T}^{\el},$ can be calculated from equations \eqref{CSincEL} and \eqref{SIGS1}. We note that the shear stress $T_{12}(t)$ only depends on the relaxation function $R_5(t)$ as in the linear regime (see equation \eqref{linTshear}). In the small-strain limit, $T_{11}$ and $T_{22}$ in \eqref{sigmatsfinaladim} tend to zero and the equation for $T_{12}$ becomes identically equal to \eqref{linTshear}.

\subsection{Anti-plane (longitudinal) shear}

\begin{figure}[b!]
\centering
\subfigure[Undeformed block]{
	\centering
	\includegraphics[scale=0.6]{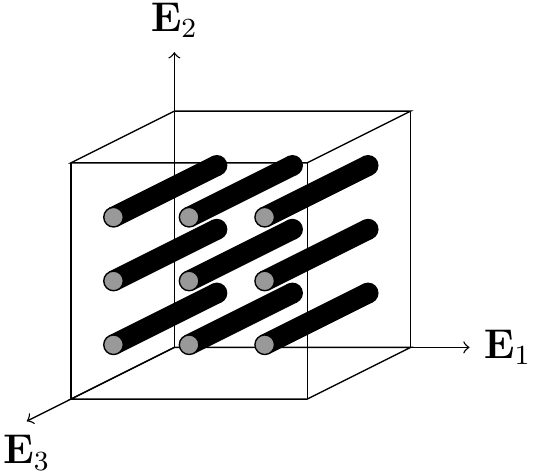}}
\qquad
\subfigure[Longitudinal shear]{
	\centering
	\raisebox{18pt}{
	\includegraphics[scale=0.6]{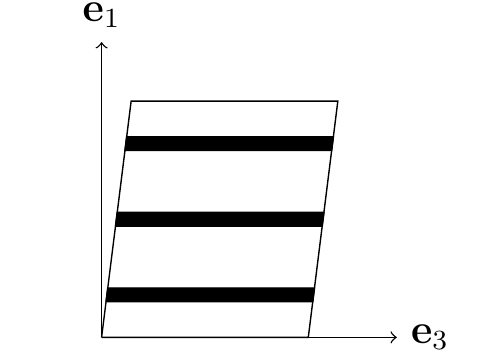}}}
\caption{A block of a TI material with fibres pointing in the direction of the $\ee_3$-axis under the simple transverse shear deformation in \eqref{ls}.}\label{fig:ls}
\end{figure}

Finally, let us consider \textit{longitudinal shear} - a simple shear along the fibre direction as depicted in Figure \ref{fig:ls}. This deformation can be written in the following form:
\begin{align}\label{ls}
x_1(t)&=X_1,  & x_2(t)&=X_2, & x_3(t) &= \kappa_3(t)X_1+X_3,
\end{align}
where
$\kappa_3(t)$ is the amount of shear in the $\ee_1$-$\ee_3$ plane. The deformation gradient and the left Cauchy-Green strain tensors associated with this anti-plane shear deformation are
\begin{equation}\label{FBls}
\ff(t)=\left(
\begin{array}{ccc}
1 &0 &0\\
0 &1 &0\\
\kappa_3(t) &0 &1
\end{array}\right)
\qquad \text{and}\qquad
\B(t)=\left(
\begin{array}{ccc}
1 &0 &\kappa_3(t)\\
0 &1 &0\\
\kappa_3(t) &0 &1+\kappa_3(t)^2
\end{array}\right).
\end{equation}
A traction-free boundary condition is imposed on the lateral surface with normal $\textbf{N}=\{0,1,0\}$ in the undeformed configuration, which leads to
\begin{equation}
T_{12}=T_{23}=T_{22}=0, \qquad \qquad \forall t.
\end{equation}
Therefore, the non-zero components of the Cauchy stress in \eqref{VEsigmaINC2} are
\begin{equation}\label{sigmals}
\begin{split}
T_{11}(t)&=-p(t)+\dfrac{T_{11}^{\el}(t)-T_{22}^{\el}(t)}{2}
+\int^t_0\dfrac{R_5'(t-\tau)}{2\mu_T}\Big(\Pi_{\T11}^{\el}(\tau)-\Pi_{\cc11}^{\el}(\tau)\Big)\text{d}\tau\\
T_{33}(t)&=-p(t)+\tilde{T}^{\el}(t)+\int_0^t\dfrac{\mathcal{R}'(t-\tau)}{E_L} \Pi_{\Lo33}^{\el}(\tau)\text{d}\tau\\
&+ \int_0^t\dfrac{R_5'(t-\tau)}{2\mu_T}\Big( \Pi_{\T33}^{\el}(\tau)-\Pi_{\cc33}^{\el}(\tau)+2 \kappa_3 (t)\left( \Pi_{\T13}^{\el}(\tau)-\Pi_{\cc13}^{\el}(\tau)\right)
+ \kappa_3^2(t) \left(\Pi_{\T11}^{\el}(\tau)-\Pi_{\cc11}^{\el}(\tau)\right)\Big)\text{d}\tau\\
&+\int_0^t\dfrac{R_6'(t-\tau)}{2\mu_L} \big(2 \kappa_3 (t)\Pi_ {\Lo13}^{\el}(\tau)+\Pi_{\cc33}\big)\text{d}\tau\\
T_{13}(t)&=T^{\el}_{13}(t)
+\int_0^t\dfrac{R_5'(t-\tau)}{2\mu_T} \left( \Pi_{\T13}^{\el}(\tau)-\Pi_{\cc13}^{\el}(\tau)+ \left(\Pi_{\T11}^{\el}(\tau)-\Pi_{\cc11}^{\el}(\tau)\right) \kappa_3 (t)\right)\text{d}\tau
+\int_0^t\dfrac{R_6'(t-\tau)}{2\mu_L}\Pi_{\Lo13}(\tau)\text{d}\tau,
\end{split}
\end{equation}
where the Lagrange multiplier p(t) can be calculated from the traction free condition $T_{22}=0$:
\begin{equation}\label{lagrls}
p(t)=- \dfrac{ T^{\el}_{11}(t) -  T^{\el}_{22}(t)}{2} +  \int^t_0 \dfrac{R_5'(t-\tau)}{2 \mu_T} \big(\Pi^{\el}_{\T22}(\tau) -\Pi^{\el}_{\cc22}(\tau)\big)\text{d}\tau.
\end{equation}
The terms
\begin{equation}\label{Sigmals}
\begin{split}
&\Pi^{\el}_{\cc11}=\Pi^{\el}_{\cc22}=\dfrac{\mu_T}{E_L}\tilde{T},\qquad \Pi^{\el}_{\cc13}=-\kappa_3\dfrac{\mu_T}{E_L}\tilde{T}\qquad\Pi^{\el}_{\cc33}= (\kappa_3^2+1)\dfrac{\mu_T}{E_L}\tilde{T},\\
&\Pi^{\el}_{\T11}=\nicefrac{1}{2}\,(T^{\el}_{11} -T^{\el}_{22}),\qquad
\Pi^{\el}_{\T22}=-\Pi^{\el}_{\T11},\qquad
\Pi^{\el}_{\T33}=\kappa_3^2\,\Pi^{\el}_{\T11},\\
&\Pi^{\el}_{\T13}=- \kappa_3\,\Pi^{\el}_{\T11},\qquad
\Pi^{\el}_{\Lo33}=\tilde{T}^{\el},
\qquad
\Pi^{\el}_{\A33}=-2\,\kappa_3\,T^{\el}_{13},
\hspace{2.8em}
\Pi^{\el}_{\A13}=T^{\el}_{13},
\end{split}
\end{equation}
along with \eqref{lagrls}, can be substituted into \eqref{sigmals} to obtain the three stress components:
\begin{equation}\label{sigmalsfinal}
\begin{split}
T_{11}(t)&=T^{\el}_{11}(t)-T^{\el}_{22}(t)
+\int^t_0\dfrac{R_5'(t-\tau)}{2\mu_T}\left(T^{\el}_{11}(\tau)-T^{\el}_{22}(\tau)\right)\text{d}\tau,\\
T_{33}(t)&=T_{33}^{\el}(t)-T_{22}^{\el}(t)+\int^t_0\dfrac{\mathcal{R}'(t-\tau)}{E_L}\tilde{T}^{\el}(\tau)\text{d}\tau
-2\int^t_0\dfrac{R_6'(t-\tau)}{2\mu_L}T^{\el}_{13}(\tau) (\kappa_3(\tau) - \kappa_3(t))\\
&+\int^t_0\dfrac{R_5'(t-\tau)}{2\mu_T}\Big(\dfrac{T^{\el}_{11}(\tau) - T^{\el}_{22}(\tau)}{2} 
+ \left(\kappa(\tau) -\kappa (t)\right)^2 \left(\dfrac{T_{11}^{\el}(\tau)-T_{22}^{\el}(\tau)}{2}-\dfrac{\mu_T}{E_L}\tilde{T}^{\el}(\tau)\right)\Big)\text{d}\tau,\\
T_{13}(t)&=T^{\el}_{13}(t)
+\int^t_0\dfrac{R_6'(t-\tau)}{2\mu_L}T^{\el}_{13}(\tau)\text{d}\tau
-\int^t_0\dfrac{R_5'(t-\tau)}{2\mu_T}\left(\dfrac{T^{\el}_{11}(\tau) - T^{\el}_{22}(\tau)}{2}-\dfrac{\mu_T}{E_L}\tilde{T}^{\el}(\tau)\right) \left(\kappa_3(\tau)-\kappa_3(t)\right) \text{d}\tau.
\end{split}
\end{equation}
The shear stress in \eqref{sigmalsfinal} depends on two relaxation functions, unlike its linear counterpart in \eqref{linLshear}, which depends only on $R_6(t)$; however, in the small-strain limit, $T_{11}$ and $T_{33}$ in \eqref{sigmalsfinal} tend to zero and the equation for $T_{13}$ becomes identically equal to \eqref{linLshear}. In the next section we illustrate some key properties of the proposed model.

\section{Key features of the modified TI QLV model}\label{sec:res}
A common procedure for investigating the time-dependent behaviour of a \begin{figure}[b!]
\centering
\subfigure[]{\includegraphics[scale=0.4]{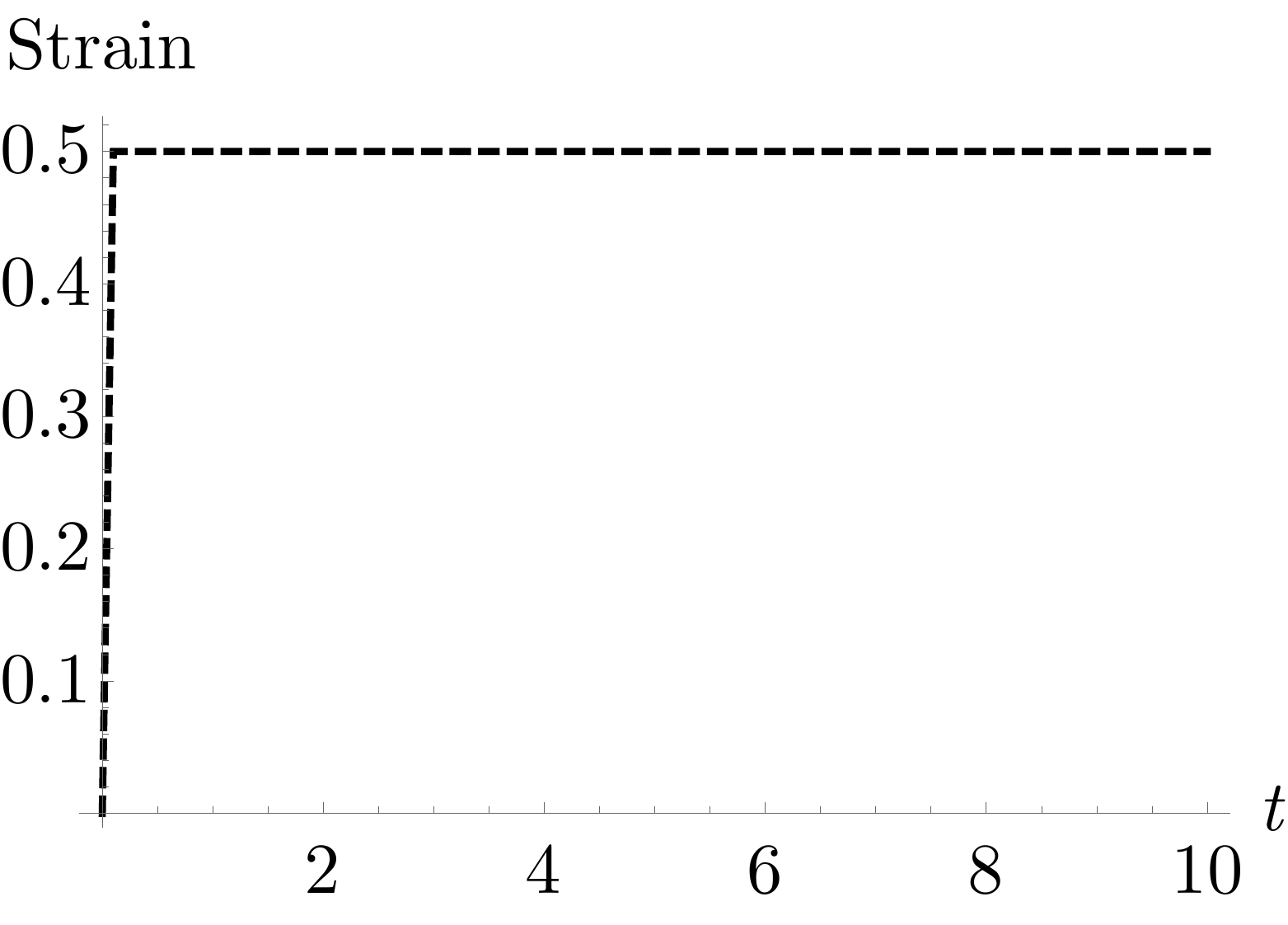}}\hspace{3em}
\subfigure[]{\includegraphics[scale=0.4]{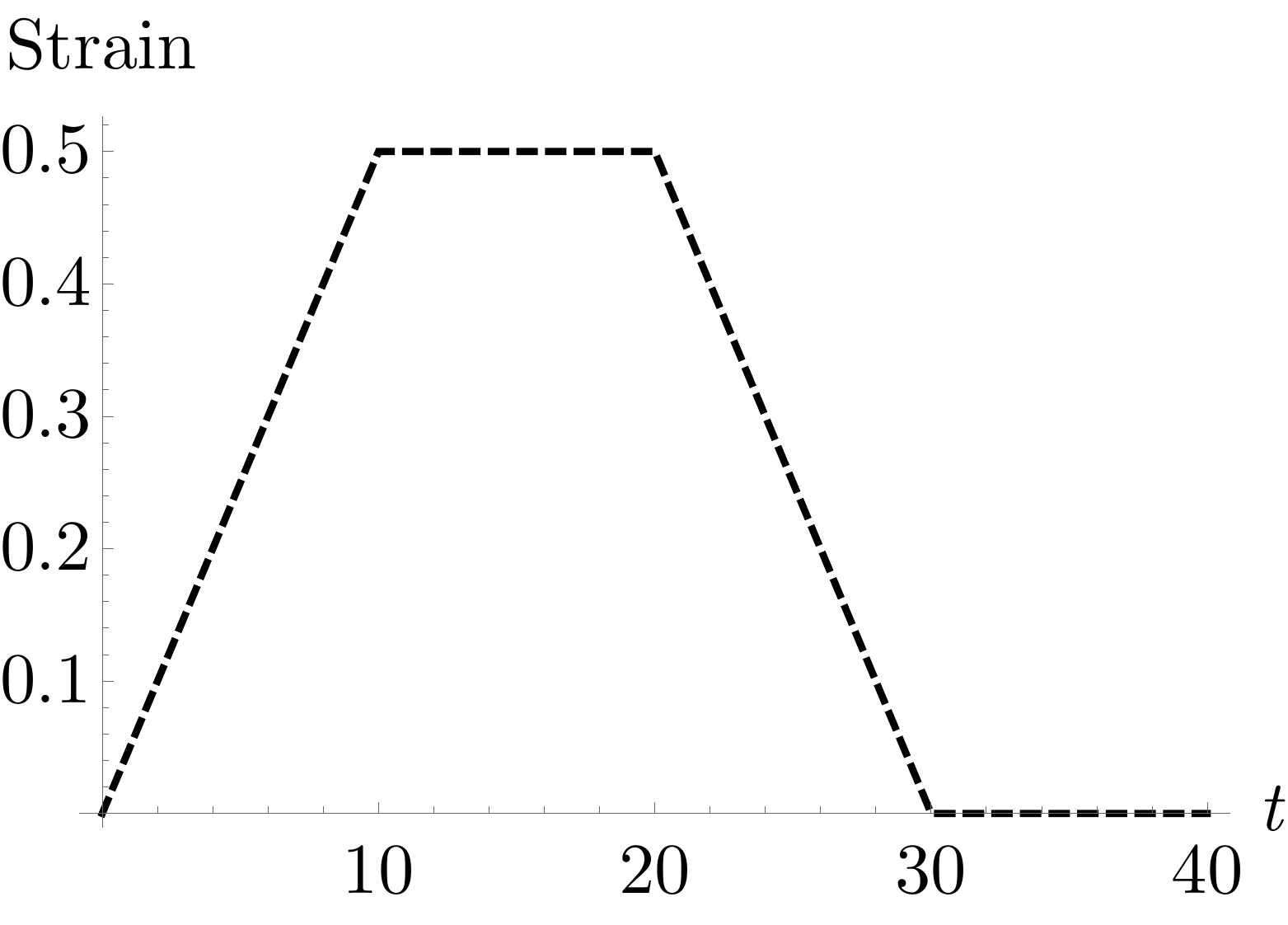}}\caption{Input time-dependence of the strain in (a) a step-and-hold stress relaxation test, and (b) a ramp-and-hold relaxation test. Note that the strain in (a) is often modelled by a Heaviside function; however, in practice, a step-change in strain cannot be achieved experimentally as this would require an infinite strain-rate; therefore, instead, a very rapid strain rate is used, as illustrated.}\label{fig:test}
\end{figure} material is to perform a \textit{step-and-hold} test. Usually, either the deformation or the load can be imposed on a sample of the material being tested. In the former case, the test is called a stress relaxation test, in the latter, a creep test. In this section, we shall focus on stress relaxation tests. In a stress relaxation test, the strain in the sample is increased very rapidly up to a maximum value, and is then held fixed during the holding phase. The response of the material is generally recorded in terms of forces or moments. Another type of test is the \textit{ramp-and-hold} test, in which the sample is deformed over a finite time interval and is then, as with the step-and-hold test, held fixed for a prescribed time period. Some experiments also account for a recovery phase - a phase where the sample is allowed to return to its original state. Examples of these two types of tests are illustrated in Figure \ref{fig:test}. The step-and-hold test is impossible to achieve experimentally because the time-dependence of the strain is required to have the form of a Heaviside function; however, this test is very useful for theoretical purposes, especially for comparing the stress relaxation responses of different models. The ramp-and-hold test provides additional information when one is interested in studying the recovery behaviour of the sample as well as its relaxation behaviour. In the next section, these two types of test will be used to illustrate the main features of the proposed modified TI QLV model.

\subsection{Comparison between linear viscoelasticity and modified TI QLV}
We first show that the modified QLV model proposed in \eqref{VEsigmaINC2} is equivalent to the linear model \eqref{siLINTIINCOMP} in the small-strain regime for the three modes of deformation described in Section \ref{sec:def}. To proceed, it is necessary to choose a specific form for the strain energy function $W$. We shall choose $W$ to be of the form:
\begin{equation}\label{W}
W=W_{\iso}+\dfrac{\mu_T-\mu_L}{2} (2 I_4-I_5-1) +\dfrac{E_L+\mu_T-4 \mu_L}{16} (I_4-1)(I_5-1),
\end{equation}
where
\begin{equation}\label{Wiso}
W_{\iso}=\dfrac{\mu_T}{2}  \big(\alpha_{\MR}  (I_1-3)+(1-\alpha_{\MR}) (I_2-3)\big),\qquad \text{with}\quad \alpha_{\MR}\in[0,1].
\end{equation}
This expression was proposed for modelling the nonlinear behaviour of TI incompressible soft tissues \cite{MURPHY201390} and it is consistent with the linear theory in the small strain limit. We note that the isotropic contribution to the strain energy takes the form of a Mooney-Rivlin function, with non-dimensional parameter $\alpha_{\MR}$. We further assume that the reduced relaxation functions take the form of classical one-term Prony series:
\begin{equation}\label{prony}
\begin{split}
&\dfrac{\mathcal{R}(t)}{E_L}=\dfrac{E_{L \infty}}{E_L}+\left(1-\dfrac{E_{L \infty}}{E_L}\right)e^{-\nicefrac{t}{\tau_{\mathcal{R}}}},\quad \dfrac{R_5(t)}{2 \mu_T}=\dfrac{\mu_{T \infty}}{\mu_T}+\left(1-\dfrac{\mu_{T \infty}}{\mu_T}\right)e^{-\nicefrac{t}{\tau_5}}\\
& \text{and} \quad \dfrac{R_6(t)}{2 \mu_L}=\dfrac{\mu_{L \infty}}{\mu_L}+\left(1-\dfrac{\mu_{L \infty}}{\mu_L}\right)e^{-\nicefrac{t}{\tau_6}},
\end{split}
\end{equation}
with $E_{L\infty}$, $\mu_{T\infty}$ and $\mu_{L\infty}$ indicating the long-time analogues of $E_L$, $\mu_T$ and $\mu_L$, respectively, and $\tau_{\mathcal{R}}$, $\tau_5$ and $\tau_6$ being the associated relaxation times. We note that, by assuming that $\mathcal{R}(t)$ takes the form of a one-term Prony series, the relaxation function $R_4(t)=\mathcal{R}(t)+\nicefrac{1}{2}R_5(t)-2R_6(t)$ will, in general, be a three-term Prony series. Alternatively, we could have chosen $R_4(t)$ to be a one-term Prony series, which would have led $\mathcal{R}(t)$ to be a three-term series; however, none of the results that follow would have changed qualitatively had we instead chosen to make that assumption.

\begin{figure}[t!]
\centering
\subfigure[Uniaxial extension]{\includegraphics[scale=0.33]{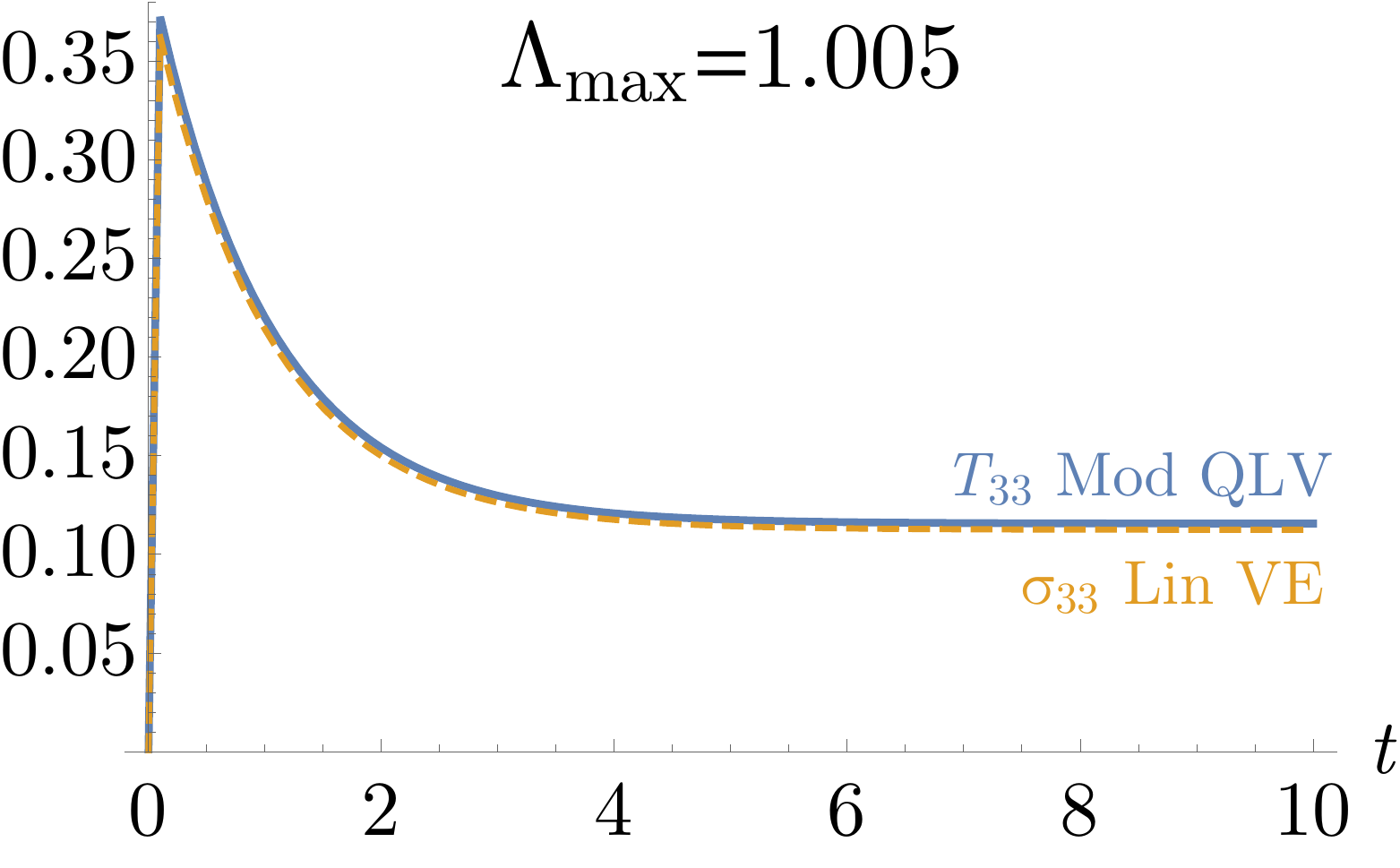}}
\subfigure[Transverse shear]{\includegraphics[scale=0.33]{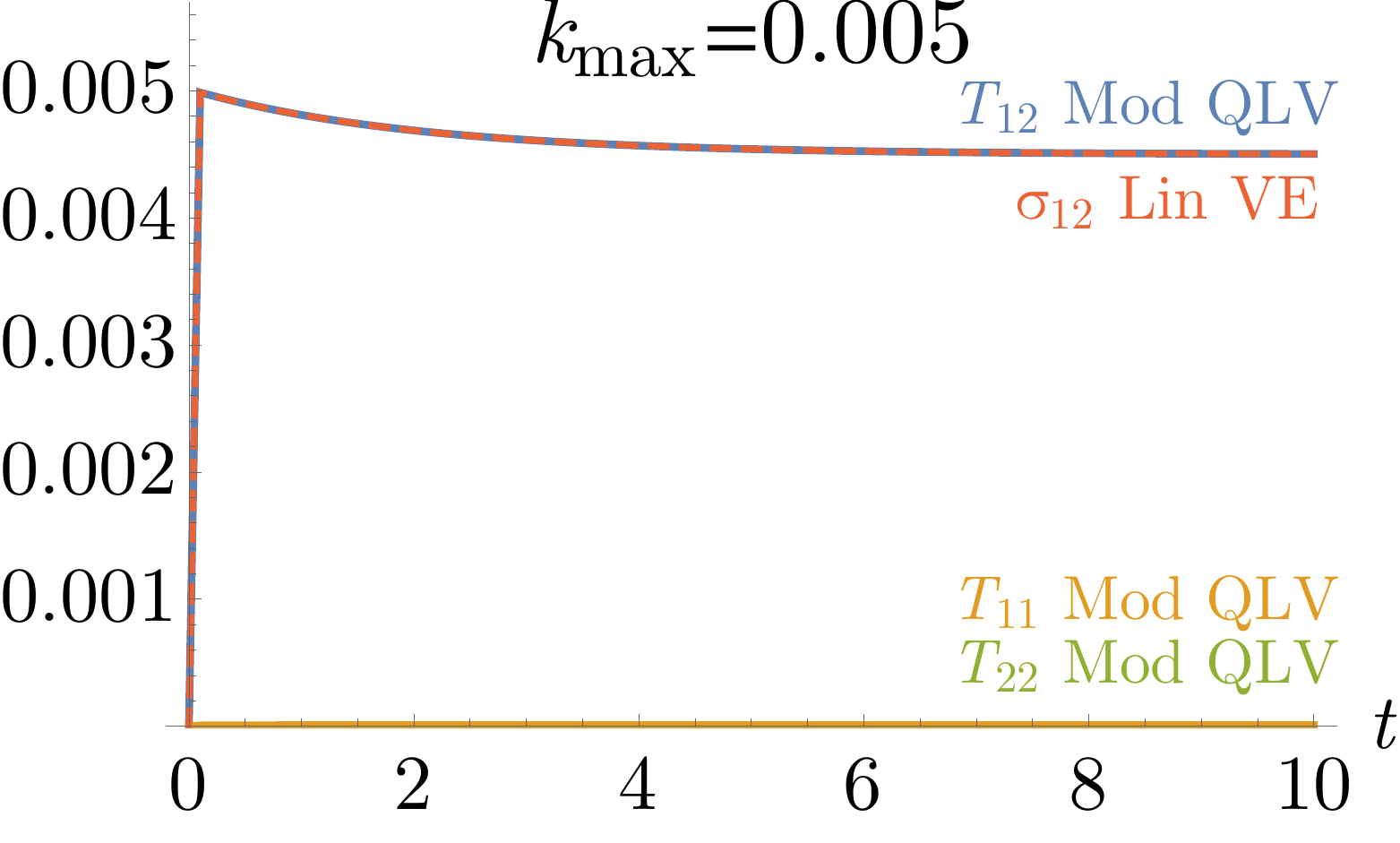}}
\subfigure[Longitudinal shear]{\includegraphics[scale=0.33]{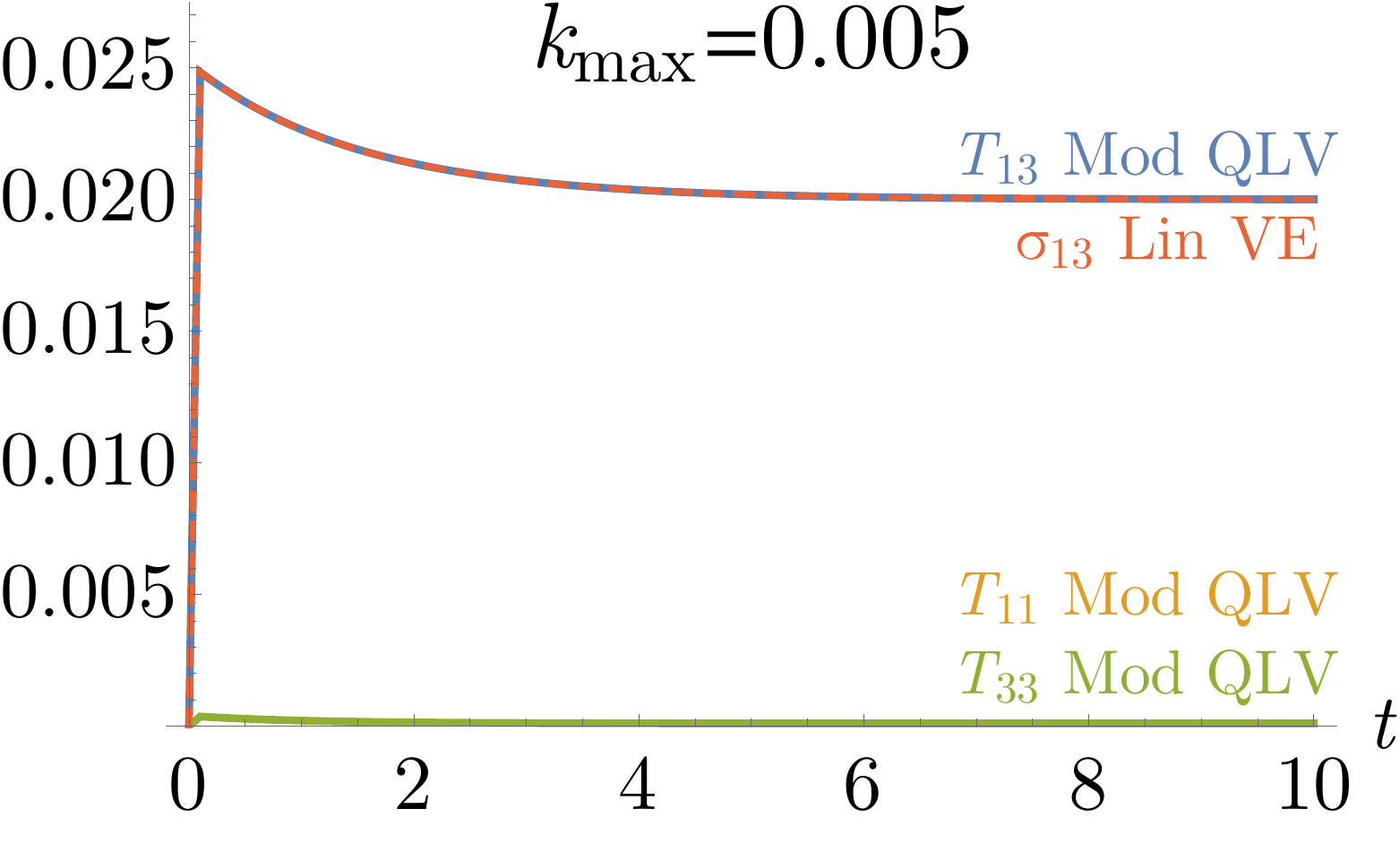}}
\subfigure[Uniaxial extension]{\includegraphics[scale=0.33]{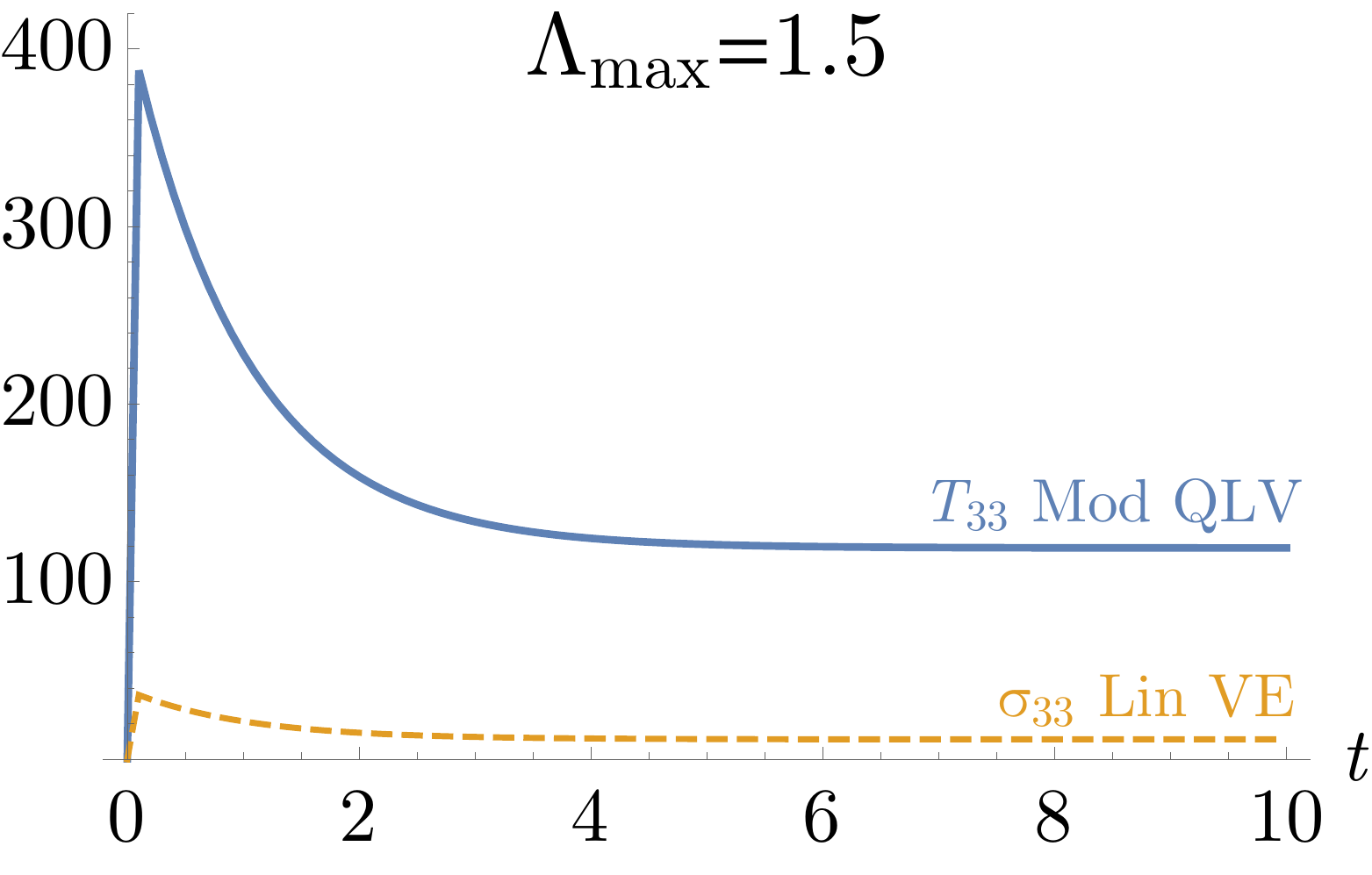}}
\subfigure[Transverse shear]{\includegraphics[scale=0.33]{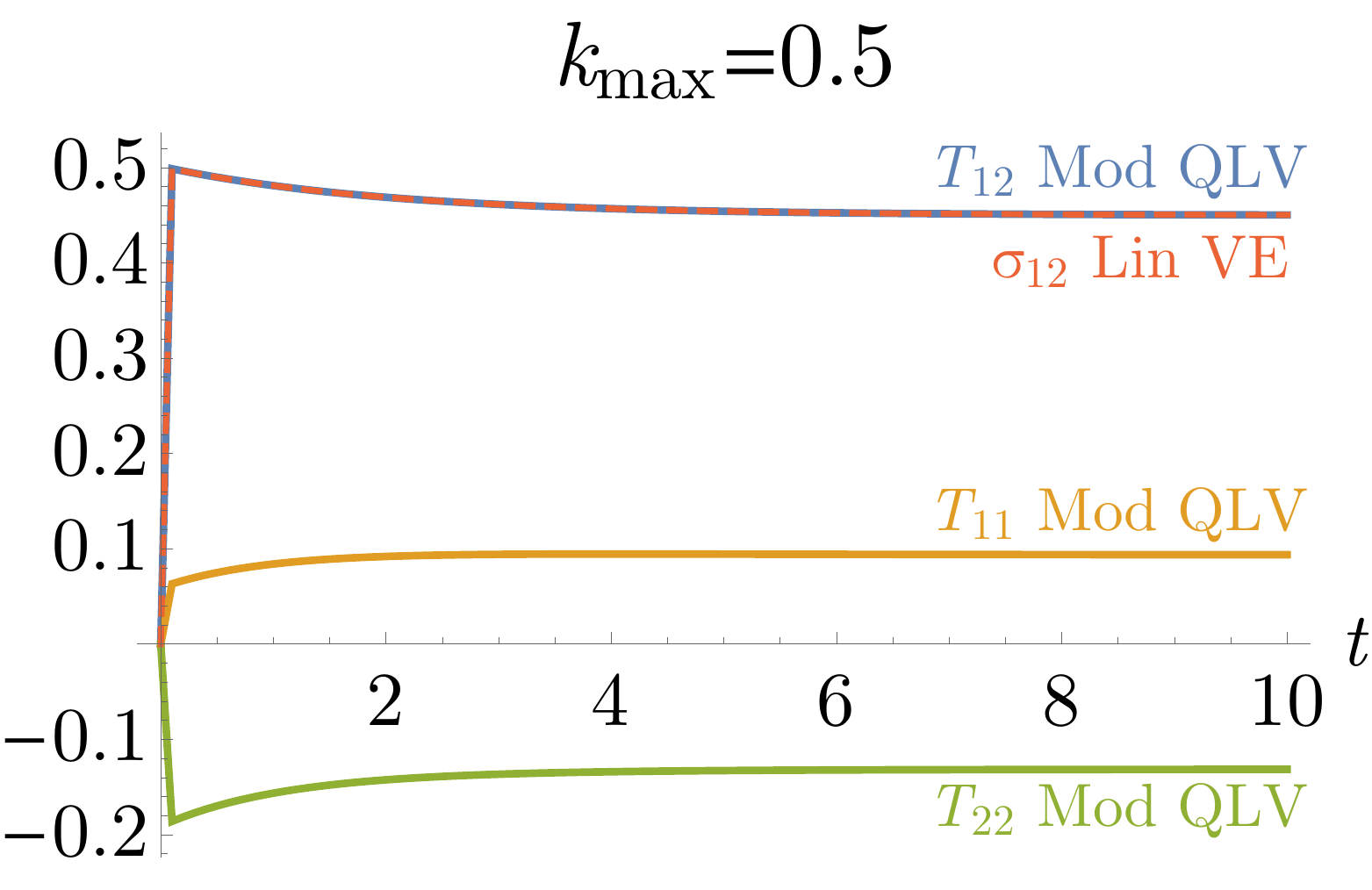}}
\subfigure[Longitudinal shear]{\includegraphics[scale=0.33]{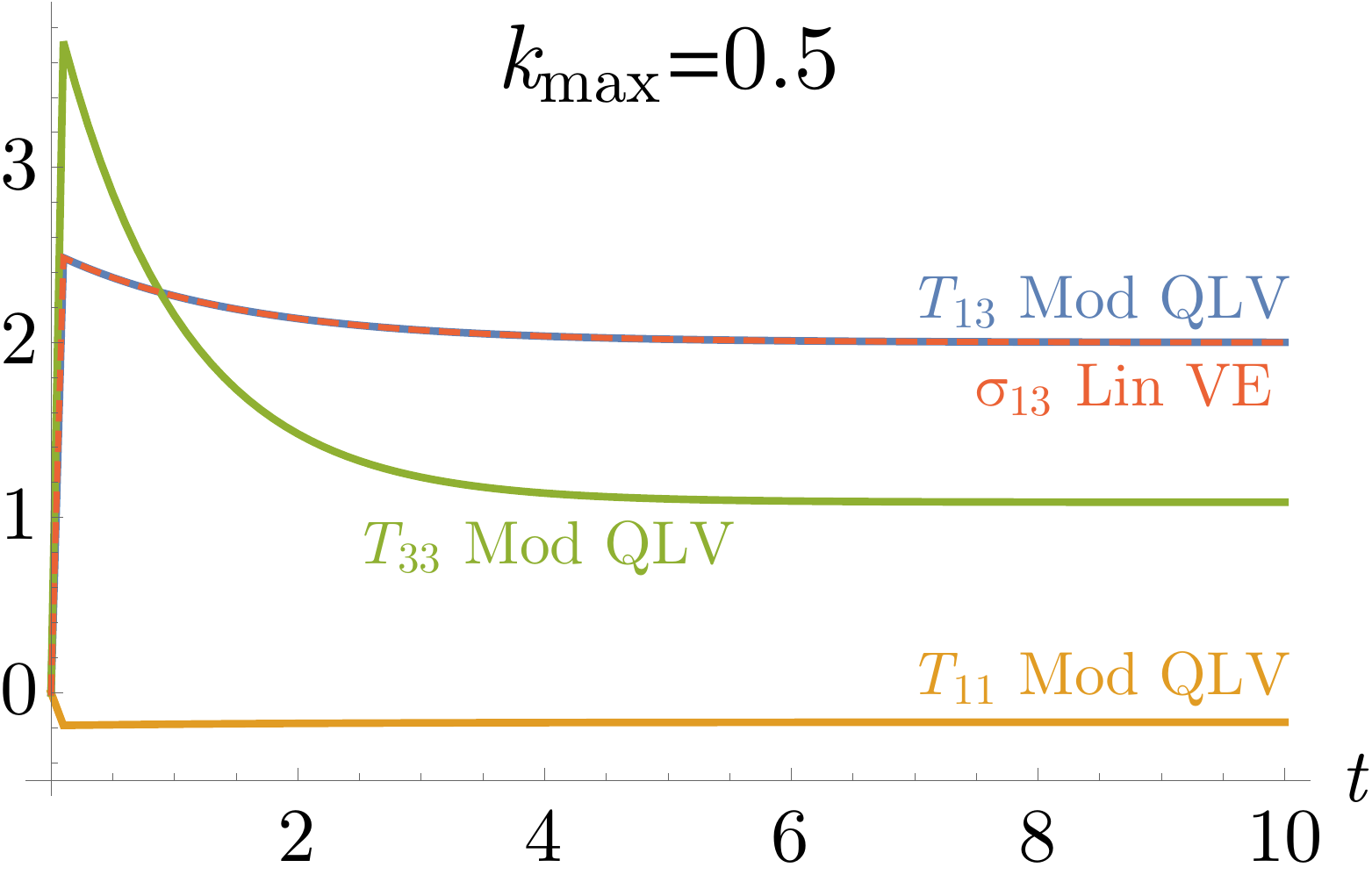}}\caption{Comparison between the linear TI model (Lin VE) \eqref{siLINTIINCOMP} and the prososed modified TI QLV model (Mod QLV) \eqref{VEsigmaINC2} of the resulting stresses induced for the three modes of deformation: uni-axial extension, transverse shear and longitudinal shear. Stress responses to step-and-hold tests in the small-strain regime are plotted in (a), (b) and (c) and in the large-strain regime in (d), (e) and (f). The curves are obtained by setting: $\nicefrac{E_{L\infty}}{E_L}=0.3$, $\tau_{\mathcal{R}}=1$, $\nicefrac{\mu_{T\infty}}{\mu_T}=0.9$ , $\tau_5=2$, $\nicefrac{\mu_{L\infty}}{\mu_L}=0.8$ , $\tau_6=1.5$, $E_L/\mu_T=75$ and $\mu_L/\mu_T=5$. All stresses shown are normalised by $\mu_T$.}\label{fig:linVSqlv}
\end{figure}

We now consider the three modes of deformation illustrated in the previous section (uni-axial extension, transverse shear and longitudinal shear) and compare the stress responses predicted by the proposed model to those of its linear counterpart derived in Section \ref{EXPTS}. We consider a step-and-hold test where the strain consists of a rapid ramp (of $0.1\text{s}$) followed by a holding phase (of $9.99\text{s}$), as depicted in Figure \ref{fig:test}(a). Figure \ref{fig:linVSqlv} shows the results in both the small (0.5\%)-strain and the large (50\%)-strain regimes. We note that in the small-strain regime the predictions of the two models are in agreement in all the three modes of deformation, thus, the modified TI QLV model is able to recover the linear limit correctly. Indeed, the shear stresses $\sigma_{12}(t)$ and $\sigma_{13}(t)$ in \eqref{linTshear} and \eqref{linLshear}, respectively, are recovered by taking the limit $\textbf{T}^{\el}\rightarrow\boldsymbol{\sigma}^{\el}$. Moreover, the stress $\sigma_{33}(t)$ in \eqref{linext} can be obtained by considering $\Lambda\rightarrow1+\epsilon_{33}$. In the large-strain regime the results for the two models vary considerably. In particular, by comparing Figures \ref{fig:linVSqlv}(b) and \ref{fig:linVSqlv}(c) with \ref{fig:linVSqlv}(e) and \ref{fig:linVSqlv}(f), we observe that, although the shear stress responses are the same for both models, there is a large discrepancy for the normal stresses. The linear TI model always predict zero normal stresses for all $t$ both in the small- and the large-strain regimes, whereas the modified TI QLV predicts non-zero normal stresses. This feature of the modified TI QLV model will be further analysed in the following sections.

\subsection{Comparison between modified isotropic QLV and modified TI QLV}

We now consider a step-and-hold test in transverse shear and calculate the stress response predicted by the proposed model. We then implement the model proposed by De Pascalis \textit{et al.} \cite{de2014nonlinear}, where a single relaxation function is used to account for the behaviour of an incompressible, isotropic material, and compare the two results in order to highlight the main differences between the models.

\begin{figure}[t]
\centering
\subfigure{\includegraphics[scale=0.4]{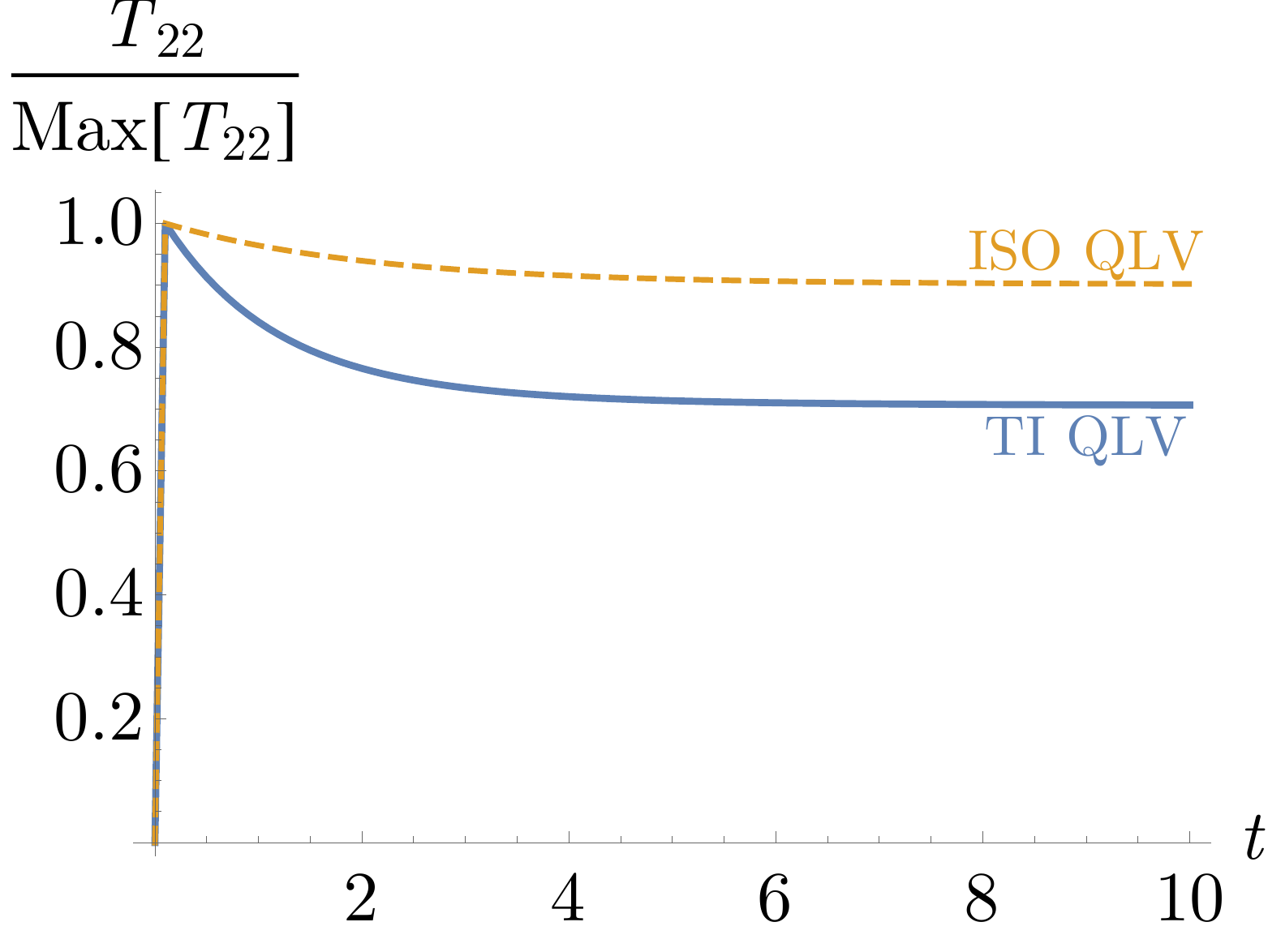}}\hspace{3em}
\subfigure{\includegraphics[scale=0.4]{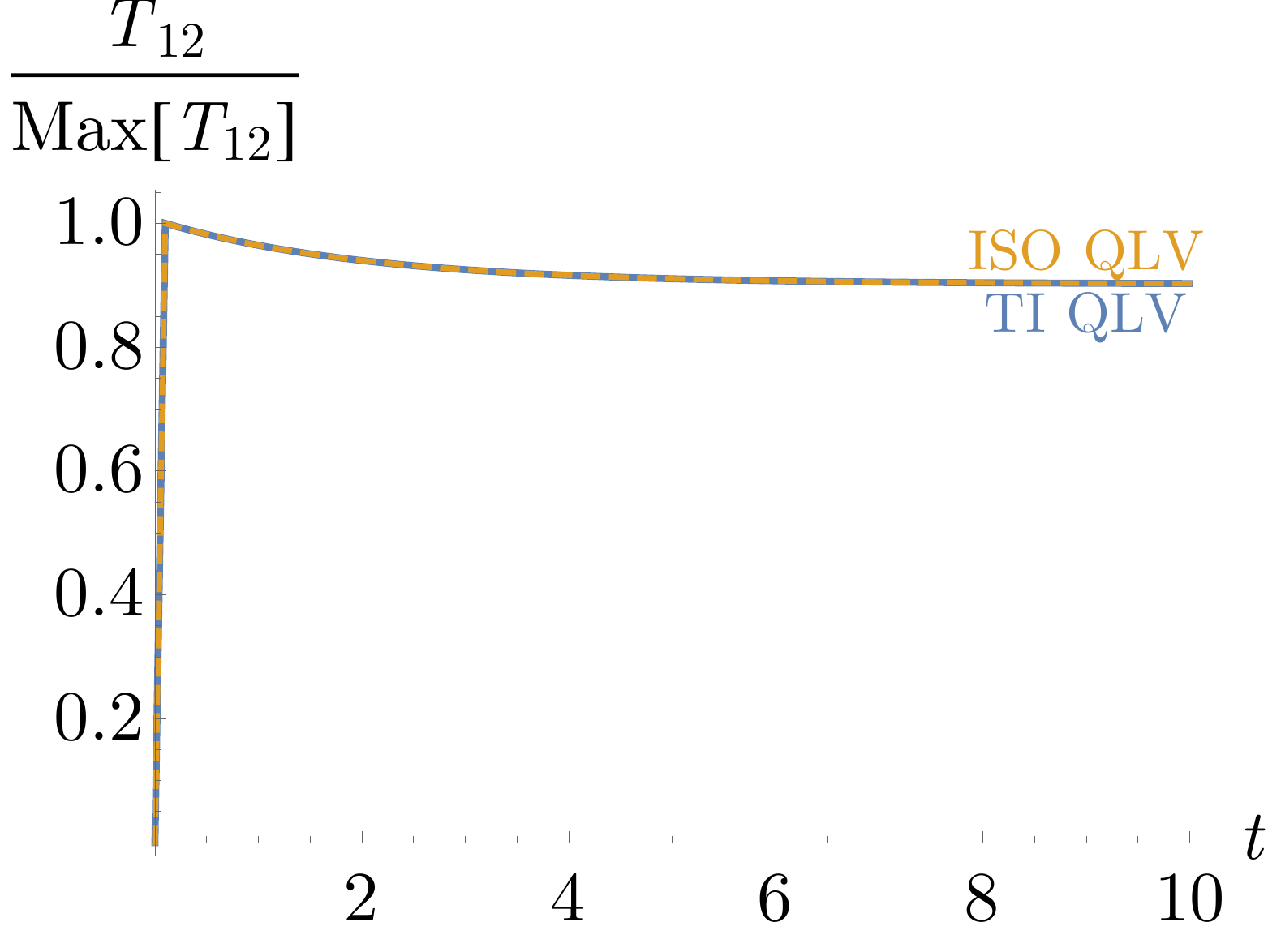}}
\caption{The predicitons of the proposed modified TI QLV model (blue) and the modified isotropic QLV model from \cite{de2014nonlinear} (yellow) for a material under transverse shear. The blue curves have been obtained by setting $\alpha_{\MR}=0.25$, $\nicefrac{E_{L\infty}}{E_L}=0.3$, $\tau_{\mathcal{R}}=1$, $\nicefrac{\mu_{T\infty}}{\mu_T}=0.9$ , $\tau_5=2$ and $E_L/\mu_T=75$, the yellow curves by setting $\nicefrac{E_{L\infty}}{E_L}=\nicefrac{\mu_{T\infty}}{\mu_T}=0.9$, $\tau_{\mathcal{R}}=\tau_5=2$ and $E_L/\mu_T=3$. The latter equation is a result of the assumption that the isotropic material is incompressible.}\label{TIvsISO}
\end{figure}

We consider the strain to consist of a rapid ramp (of $0.1\text{s}$) where the amount of shear $\kappa$ increases up to $0.5$ followed by a holding phase (of $9.99\text{s}$), as depicted in Figure \ref{fig:test}(a). We then calculate the normal ($T_{22}$) and shear ($T_{12}$) stress components from equation \eqref{sigmatsfinaladim}. In Figure \ref{TIvsISO}, the results for the two models are presented. The stress components are normalised on their respective maximum values at $0.1\text{s}$. We recall that the modified isotropic QLV model \cite{de2014nonlinear} can be recovered from the proposed model by setting $R_2(t)=R_3(t)=R_4(t)=0$, $R_6(t)=R_5(t)$ $\forall t$, $\alpha=\beta=0$ and $\mu_L=\mu_T$. For the modified TI QLV model we have set $\nicefrac{E_L\infty}{E_L}\ll\nicefrac{\mu_T\infty}{\mu_T}$ so that the function $\mathcal{R}(t)$ relaxes considerably more than $R_5(t)$.

Although the predictions of the two models are in agreement for the shear stress response, their normal stress responses differ. For the modified TI QLV model, the relaxation curve of $T_{22}$ is mostly determined by the integrals associated with $\mathcal{R}(t)$, whereas, for the modified isotropic QLV model, the relaxation behaviour is entirely dictated by the function $R_5(t)$, as shown in Figure \ref{TIvsISO}, and the normalised normal and shear stress relaxation behaviours are identical. This is an important and unique property displayed by the proposed model and it arises from the presence of a tensorial relaxation function with distinct components. In general, the stress response predicted by the proposed modified TI QLV model will depend on the competition between the integrals associated with each of the three relaxation functions $\mathcal{R}(t)$, $R_5(t)$ and $R_6(t)$. The modified isotropic QLV model lacks this property and so do all QLV models that incorporate a single scalar relaxation function. It is therefore of utmost importance to include more than a single relaxation function when modelling TI materials that exhibit direction-dependent relaxation behaviours. Otherwise, the risk of running into errors when measuring the mechanical parameters can dramatically increase.

In the next section, we will show that the modified TI QLV model predicts the so-called Poynting effect. This is a commonly observed phenomenon in nonlinear elastic materials undergoing simple shear or torsion. Such materials tend to expand or contract in the direction perpendicular to the shear direction. To prevent such an expansion or contraction, a normal stress is required. For the transverse shear deformation illustrated in Section \ref{sec:ts}, for instance, $T_{22}$ can be either positive or negative when the material is sheared. The general convention is that a negative (positive) normal stress is associated with the positive (negative) Poynting effect.

\subsection{The Poynting effect}\label{sec:PE}
It is well known that nonlinear elastic materials display the Poynting effect. Since 1909, when the phenomenon was first discovered by Poynting \cite{poynting1909pressure}, many studies have focused on theoretical and experimental aspects of this effect. In particular, within the class of incompressible, hyperelastic materials under simple shear, the problem has been studied for isotropic \cite{mihai2011positive,mihai2013numerical} and TI media \cite{destrade2015dominant,horgan2017poynting}. Specifically, for isotropic materials, it has been shown that no Poynting effect occurs in neo-Hookean materials. Similarly, fibre-reinforced neo-Hookean materials exhibit no Poynting effect in transverse shear. Interestingly, however, the Poynting effect in viscoelastic materials seems to have received little attention. This section highlights how the proposed modified TI QLV model is capable of giving new theoretical insights into the Poynting effect in viscoelastic TI materials.

\begin{figure}[t]
\begin{center}
\includegraphics[scale=0.4]{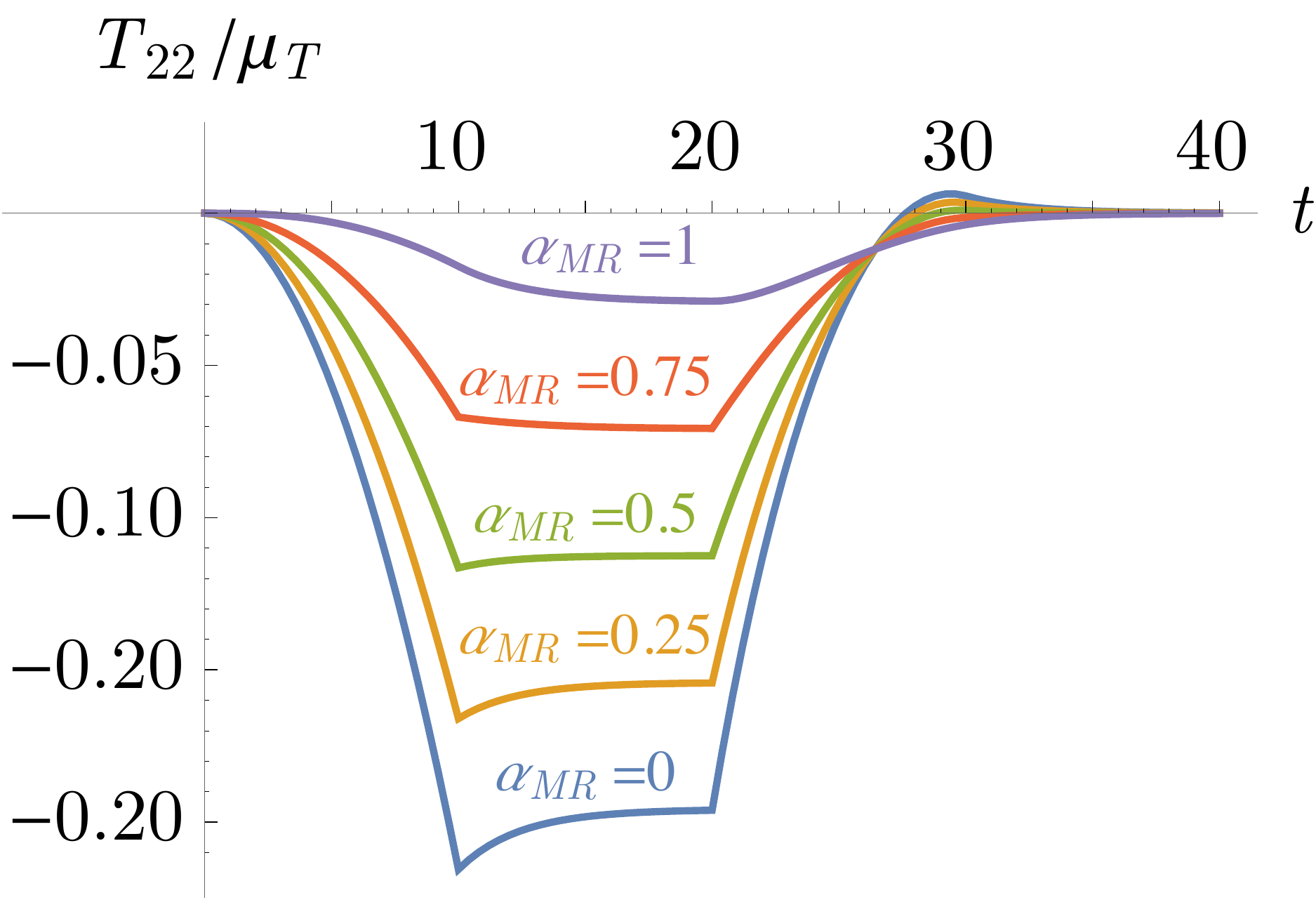}
\end{center}
    \caption{The normal stress response of the modified TI QLV material under transverse shear. The normal stress $\nicefrac{T_{22}}{\mu_T}$ is plotted against the time $t$ for different values of $\alpha_{\MR}=\{0,0.25,0.5,0.75,1\}$, for fixed $\nicefrac{E_{L\infty}}{E_L}=0.66, \nicefrac{\mu_{T\infty}}{\mu_T}=0.9$, $\tau_{\mathcal{R}}=2.5$ and $\tau_5=2$.}\label{fig:ts}
\end{figure}

We now consider a ramp-and-hold test as in Figure \ref{fig:test}(b) for the transverse shear deformation in equation \eqref{ts}. The amount of shear $\kappa$ increases up to 0.5 in $10\text{s}$, is then held constant for $10\text{s}$ before decreasing back to $0$ in $10\text{s}$ and being held there for a further 10s. We calculate the corresponding normal stress response $T_{22}$ from \eqref{sigmatsfinaladim}. Upon normalising the stress on $\mu_T$, the parameters appearing in equation \eqref{sigmatsfinaladim} are the mechanical parameters $\nicefrac{E_{L\infty}}{E_L}$, $\nicefrac{\mu_{T\infty}}{\mu_T}$ and $\alpha_{\MR}$, and the relaxation times, $\tau_{\mathcal{R}}$ and $\tau_5$. In Figure \ref{fig:ts}, we plot curves for the stress $T_{22}/\mu_T$ for different values of $\alpha_{\MR}$. The Poynting effect decreases with increasing $\alpha_{\MR}$, but is still non-zero even for $\alpha_{\MR}=1$. We recall that $\alpha_{\MR}=1$ is associated with a neo-Hookean TI material, whereas $\alpha_{\MR}=0$ indicates a pure Mooney-Rivlin TI material.

\begin{figure}[t]
\centering
\includegraphics[scale=0.48]{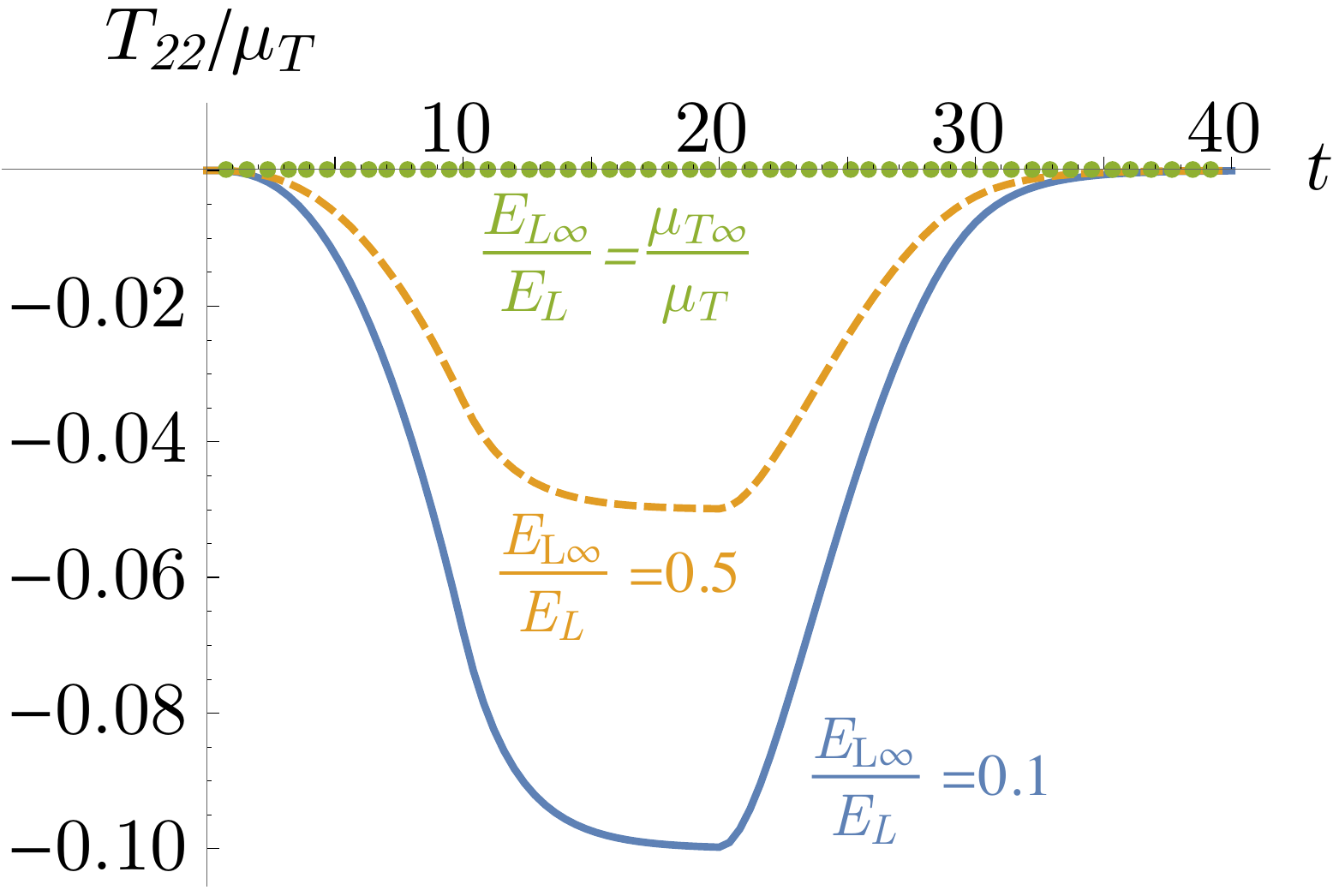}
\caption{The Poynting effect predicted by the modified TI QLV model. The normal stress component $\nicefrac{T_{22}}{\mu_T}$ is plotted against time $t$ for a neo-Hookean viscoelastic TI material ($\alpha_{\MR}=1$). The following parameters have been set: $\mu_{T\infty}/\mu_T=0.9,\tau_5=\tau_{\mathcal{R}}=2$, whilst $\nicefrac{E_{L\infty}}{E_L}$ spans over \{0.1,\,0.5,\,0.9\}.}\label{fig:compet}
\end{figure}

As discussed at the beginning of the section, a hyperelastic model with a strain energy function given by \eqref{W} predicts a normal stress $T_{22}^{\el}\!=\!0$ for fibre-reinforced neo-Hookean materials under simple shear, indicating no Poynting effect. In contrast to the hyperelastic model, the modified TI QLV model \textit{does} predict a (positive) Poynting effect, as illustrated by the purple curve ($\alpha_{\MR}=1$) in Figure \ref{fig:ts}. This interesting behaviour suggested by the numerical results will of course need to be verified by bespoke experiments; however, the curves depicted in Figure \ref{fig:compet} remarkably indicate that the Poynting effect is dictated by the competition between the two relaxation functions appearing in the equation for $T_{22}$ in \eqref{sigmatsfinaladim}. We further note that when $\nicefrac{\mathcal{R}(t)}{E_L}=\nicefrac{R_5(t)}{2\mu_T}$ for all $t$, i.e. in the isotropic case, the Poynting effect vanishes (green, dotted curve) in agreement with the elastic behaviour of neo-Hookean TI materials.

\section{Conclusions}\label{sec:concl}
In this paper, a modified TI QLV theory for finite deformations has been developed. Transverse isotropy is accommodated both in terms of elastic anisotropy \textit{and} relaxation functions, thus improving on existing scalar relaxation function TI QLV models. The numerical results presented in Section \ref{sec:res} have shown that incorporating distinct relaxation functions is crucial when modelling TI materials, and that simplified models with only one relaxation function would fail to capture the Poynting effect, for example. Another appeal of the proposed model is that the relaxation functions can be determined from small-strain mechanical tests. Moreover, the formulation in terms of tensor bases motivates similar analyses for other important viscoelastic anisotropies, such as orthotropy. Finally, the theory developed here can be used as a starting point for more complex, fully three-dimensional nonlinear viscoelastic theories that are able to incorporate strain-dependent relaxation.

%
%\ethics{The research work did not involve active collection of human data or any other ethical issues.}
%
%\dataccess{This work does not have any experimental data.}
%
%\conflict{The authors have no conflicts of interest to declare.}
%
%\contri{All authors equally contributed to the study and to the draft of the manuscript. All authors gave final approval for publication.}
%
\section{Acknowledgments}
This work has received funding from the European Union's Horizon 2020 research and innovation programme under the Marie Sk\l odowska-Curie grant agreement No. 705532 (V.B.). T.S. and W.J.P. thank the Engineering and Physical Sciences Research Council for supporting this work (via grants EP/L017997/1 and EP/L018039/1).

\appendix

\section{Transversely isotropic basis tensors}   \label{app:tens}

\subsection{The Hill basis}
Fourth order TI tensors can be written down in a compact manner by defining appropriate basis tensors. A TI basis has \textit{six} basis tensors. One such basis, often associated with Hill \cite{parnell2016eshelby} is the following, introducing an axis of anisotropy in the direction of the unit vector $\mathbf{M}$ and defining $\boldsymbol{\Theta}=\mathbf{I}-\mathbf{m}\otimes\mathbf{m}$,
\begin{align}
\mathbb{H}^{1}_{ijk\ell} &= \frac{1}{2}\Theta_{ij}\Theta_{k\ell}, &
\mathbb{H}^{2}_{ijk\ell} &= \Theta_{ij}m_k m_{\ell}, &
\mathbb{H}^{3}_{ijk\ell} &= \Theta_{k\ell}m_{i}m_{j}, \label{H1}
\end{align}
\vspace{-0.7cm}
\begin{align}
\mathbb{H}^{4}_{ijk\ell} &= M_{i}m_{j}m_{k}m_{\ell}, &
\mathbb{H}^{5}_{ijk\ell} &= \frac{1}{2}(\Theta_{ik}\Theta_{\ell j}+\Theta_{i\ell}\Theta_{kj}-\Theta_{ij}\Theta_{k\ell}),
\end{align}
\vspace{-0.7cm}
\begin{align}
\mathbb{H}^{6}_{ijk\ell} &= \frac{1}{2}(\Theta_{ik}m_{\ell}m_{j}+\Theta_{i\ell}m_{k}m_{j}+\Theta_{jk}m_{\ell}m_{i}+\Theta_{j\ell}m_{k}m_{i}). \label{H6}
\end{align}
It is straightforward to show that
\begin{align}
\mathbb{I} &= \mathbb{H}^1  + \mathbb{H}^4+\mathbb{H}^5+\mathbb{H}^6 \label{IHproperty},
\end{align}
where $\mathbb{I}$ is the fourth order identity tensor. The fourth order linear elastic modulus tensor $\mathbb{C}$ relates the linear stress $\CS^{\textrm{e}}$ to the linear strain $\boldsymbol{\epsilon}$ in the form $\CS^{\textrm{e}} = \mathbb{C}\boldsymbol{\epsilon}$. If the medium is TI, then $\mathbb{C}$ can be written with respect to the Hill basis in the form
\begin{align}
\mathbb{C} &= \sum_{n=1}^6 h_n \mathbb{H}^n,
\end{align}
where $h_1=2K, h_2=h_3=\ell, h_4=r, h_5=2\mu_T, h_6=2\mu_L$. Here, $K$ is the bulk modulus in the plane of isotropy and $\mu_T$ and $\mu_L$ are the transverse (in the plane of isotropy) and longitudinal shear moduli, respectively.

\subsection{Fibre-reinforced composite basis}
A basis frequently employed in the context of fibre reinforced materials is the following \cite{Spencer1984}, which is most easily defined in terms of the Hill basis,
\begin{align}
\mathbb{J}^1 &= 2\mathbb{H}^1 + \mathbb{H}^2 + \mathbb{H}^3 + \mathbb{H}^4, & \mathbb{J}^2 &= \mathbb{H}^2 - \mathbb{H}^4, & \mathbb{J}^3 &= \mathbb{H}^3 - \mathbb{H}^4, \label{appJ1}\\
\mathbb{J}^4 &= \mathbb{H}^4, & \mathbb{J}^5 &= \mathbb{H}^5-\mathbb{H}^4+\mathbb{H}^1, & \mathbb{J}^6 &= \mathbb{H}^6 + 2\mathbb{H}^4, \label{appJ2}
\end{align}
so that
\begin{align}
\mathbb{C} &= \sum_{n=1}^6 j_n \mathbb{J}^n,
\end{align}
where $j_1=\lambda, j_2=j_3=\al, j_4=\be, j_5=2\mu_T, j_6=2\mu_L$ and
\begin{align}
K &= \lambda+\mu_T, & \ell &=\lambda+\alpha, &
r &= \lambda-2\alpha+\beta - 2\mu_T+4\mu_L,
\end{align}
and it can be seen that this is the basis employed in \eqref{TIHookea}, for example.

\subsection{A basis for the modified theory of TI QLV}
As explained in Section \ref{sec:linvisc}, after equation \eqref{FungformB}, of specific importance to the quasi-linear form of the constitutive equation, one needs a basis $\mathbb{K}^n$ that sums to the identity tensor:
\begin{align}
\sum_{n=1}^6\mathbb{K}^n = \mathbb{I}. \label{appKI}
\end{align}
Note that
\begin{align}
\sum_{n=1}^6\mathbb{H}^n &\neq \mathbb{I}, & \sum_{n=1}^6\mathbb{J}^n &\neq \mathbb{I}.
\end{align}
Let us choose a linearly independent combination of the Hill bases that gives the property \eqref{appKI}:
\begin{align}
\mathbb{K}^1 &= \mathbb{H}^2-\mathbb{H}^1, & \mathbb{K}^2 &= 2\mathbb{H}^4-\mathbb{H}^3, & \mathbb{K}^3 &= 2\mathbb{H}^1-\mathbb{H}^2, \label{basisTI1} \\
\mathbb{K}^4 &= \mathbb{H}^3-\mathbb{H}^4, & \mathbb{K}^5 &= \mathbb{H}^5, & \mathbb{K}^6 &= \mathbb{H}^6, \label{basisTI4}
\end{align}
so that by \eqref{IHproperty},
\begin{align}
\sum_{n=1}^6 \mathbb{K}^n = \mathbb{H}^1 + \mathbb{H}^4 + \mathbb{H}^5 + \mathbb{H}^6 = \mathbb{I}.
\end{align}
Note that this property is not unique to the set $\mathbb{K}^n$ since any arbitrary ``addition of zero'' across the basis tensors could achieve this; however, this choice is certainly a reasonable and viable choice on which to base a modified theory of TI QLV. The basis $\mathbb{K}^n$ decomposes a second order tensor $\boldsymbol{\si}^{\textrm{e}}$ as follows:
\begin{align}
 \boldsymbol{\si}^{\textrm{e}} &= \sum_{n=1}^6 \mathbb{K}^n:\boldsymbol{\si}^{\textrm{e}} =  \boldsymbol{\si}^{\textrm{e}}_1 + \boldsymbol{\si}^{\textrm{e}}_2 + \boldsymbol{\si}^{\textrm{e}}_3 + \boldsymbol{\si}^{\textrm{e}}_4 + \boldsymbol{\si}^{\textrm{e}}_5 + \boldsymbol{\si}^{\textrm{e}}_6,
\end{align}
where
\begin{align}
\boldsymbol{\si}^{\textrm{e}}_1 &=  \tilde{\si}^{\textrm{e}}\boldsymbol{\Theta}, &
\boldsymbol{\si}^{\textrm{e}}_2 &=  2\tilde{\si}^{\textrm{e}}\mathbf{m}\otimes\mathbf{m}, &
\boldsymbol{\si}^{\textrm{e}}_3 &= \bar{\si}^{\textrm{e}}\boldsymbol{\Theta}, &
\boldsymbol{\si}^{\textrm{e}}_4 &= \bar{\si}^{\textrm{e}}\mathbf{m}\otimes\mathbf{m}, \label{sigsplit1}
\end{align}
\begin{align}
\boldsymbol{\si}^{\textrm{e}}_5 &= \boldsymbol{\si}^{\textrm{e}}- \boldsymbol{\si}^{\textrm{e}}_m + \mathbf{m}\otimes\mathbf{m}\si^{\textrm{e}}_{\parallel}- \frac{1}{2}\boldsymbol{\Theta}\left(\textnormal{tr}\boldsymbol{\si}^{\textrm{e}}-\si^{\textrm{e}}_{\parallel}\right), &
\boldsymbol{\si}^{\textrm{e}}_6 &= \boldsymbol{\si}^{\textrm{e}}_m,
\label{sigsplit3}
\end{align}
and
\begin{align}
\tilde{\si}^{\textrm{e}} &= \si^{\textrm{e}}_{\parallel}-\frac{1}{2}(\textnormal{tr}\boldsymbol{\si}^{\textrm{e}}-\si^{\textrm{e}}_{\parallel}),
& \bar{\si}^{\textrm{e}} &= 2\left(\frac{1}{2}\textnormal{tr}\boldsymbol{\si}^{\textrm{e}}-\si^{\textrm{e}}_{\parallel}\right) = 2(\si^{\textrm{e}}_{\parallel}-\tilde{\si}^{\textrm{e}}), \label{SIGS1}  \\
\si^{\textrm{e}}_{\parallel} &= \mathbf{m}\cdot\boldsymbol{\si}^{\textrm{e}}\mathbf{m}, &
\boldsymbol{\si}^{\textrm{e}}_m &= (\boldsymbol{\si}\mathbf{m})\otimes\mathbf{m} + \mathbf{m}\otimes(\boldsymbol{\si\mathbf{m}}). \label{SIGS}
\end{align}
Finally, with the tensor decompositions \eqref{gJbasis} and \eqref{GTI}, i.e.\
\begin{align}
\mathbb{R} &= \sum_{n=1}^6 R_n \mathbb{J}^n, &  \mathbb{G} &= \sum_{n=1}^6 G_n \mathbb{K}^n,
\end{align}
one can use the decomposition for $\mathbb{G}$, i.e.\ \eqref{GTI} together with \eqref{TIHookea} in \eqref{Fungform}, and equate the resulting expression to \eqref{Fungforma}, to yield the important connections
\begin{equation}
\begin{split}
G_1&= A\,R_1+B\,\,R_2+ \dfrac{A-B}{2}\,R_5,\hspace{2.8em}
G_2= \dfrac{A}{2}\,(R_1+R_3)+\dfrac{B}{2}\,(R_2+R_4-R_5+2R_6),\\
G_3&= -C\,R_1 - D\,R_2 - \dfrac{C-D}{2}\,R_5,\qquad G_4= -C\, (R_1+R_3)-D \,(R_2+R_4-R_5+2 R_6),\\
G_5&= \frac{R_5}{2 \mu_T},\qquad G_6= \frac{R_6}{2 \mu_L}, \label{Gg3}
\end{split}
\end{equation}
where
\begin{equation}
\begin{split}
A&= \frac{1}{\Delta}(\alpha-\beta-4\mu_L), \hspace{2.5em} B= \frac{1}{\Delta}(\alpha-\lambda-2\mu_T),\\
 C &= \frac{1}{\Delta}(\beta+4\mu_L-\mu_T),  \qquad D = \frac{1}{\Delta}(\mu_T-\alpha), \label{app:A1}
\end{split}
\end{equation}
\begin{align}
\Delta &= (\alpha-\lambda-2\mu_T)(\beta+4\mu_L-\mu_T)-(\alpha-\beta-4\mu_L)(\mu_T-\alpha). \label{app:A2}
\end{align}

\subsection{The components $P^{\el}_n$}
\begin{equation}\label{bigSigma}
\begin{split}
\boldsymbol{P}^{\el}_1&= A\Big(\PPi^{\el}_1+\dfrac{\PPi^{\el}_2}{2}\Big)-C\Big(\PPi^{\el}_3+\PPi^{\el}_4\Big)=J\left(A\,\tilde{T}^{\el}-C\,\bar{T}^{\el}\right)\C^{-1},\\
\boldsymbol{P}^{\el}_2&=B\Big(\PPi^{\el}_1+\dfrac{\PPi^{\el}_2}{2}\Big)- D\Big(\PPi^{\el}_3+\PPi^{\el}_4\Big)=J\left(B\,\tilde{T}^{\el}-D\,\bar{T}^{\el}\right)\C^{-1},\\
\boldsymbol{P}^{\el}_3&=\nicefrac{A}{2}\,\PPi^{\el}_2-C\,\PPi^{\el}_4=J\left(A\,\tilde{T}^{\el}-C\,\bar{T}^{\el}\right)\textbf{M}\otimes\textbf{M},\\
\boldsymbol{P}^{\el}_4&=\nicefrac{B}{2}\,\PPi^{\el}_2- D\,\PPi^{\el}_4=J\left(B\,\tilde{T}^{\el}-D\,\bar{T}^{\el}\right)\textbf{M}\otimes\textbf{M},\\
\boldsymbol{P}^{\el}_5&=\nicefrac{A}{2}\,\PPi^{\el}_1-\nicefrac{B}{2}\left(\PPi^{\el}_1+\PPi^{\el}_2\right)-\nicefrac{C}{2}\,\PPi^{\el}_3+\nicefrac{D}{2}\left(2\PPi^{\el}_4+\PPi^{\el}_3\right)+\dfrac{\PPi^{\el}_5}{2 \mu_T}\\
&=J\dfrac{A\tilde{T}^{\el}-C\bar{T}^{\el}}{2}\left(\textbf{C}^{-1}-\textbf{M}\otimes\textbf{M}\right)-J\dfrac{B\tilde{T}^{\el}-D\bar{T}^{\el}}{2}\left(\textbf{C}^{-1}+\textbf{M}\otimes\textbf{M}\right)+\dfrac{\PPi^{\el}_5}{2 \mu_T},\\
\boldsymbol{P}^{\el}_6&=B\,\PPi^{\el}_2 - 2D\, \PPi^{\el}_4 + \dfrac{\PPi^{\el}_6}{2 \mu_L}
=2J\left(B\,\tilde{T}^{\el}-D\,\bar{T}^{\el}\right)\textbf{M}\otimes\textbf{M}+ \dfrac{\PPi^{\el}_6}{2 \mu_L},\\
\end{split}
\end{equation}
where $\tilde{T}^{\el}$ and $\bar{T}^{\el}$ are analogous to $\tilde{\si}^{\el}$ and $\bar{\si}^{\el}$ in equations \eqref{SIGS1}.

%%%%%%%%%% Insert bibliography here %%%%%%%%%%%%%%
\bibliographystyle{vancouver}
\bibliography{QLVTI_arXiv.bib}

\end{document}